\documentclass[preprint,aps,prl]{revtex4-1}
\usepackage[utf8]{inputenc}
\usepackage{epsfig}
\usepackage{subfigure}
\usepackage{latexsym}
\usepackage{amsmath}
\usepackage{psfrag}
\usepackage{fullpage}
\usepackage{graphicx}
\usepackage{amssymb}
\usepackage{color}
\usepackage{placeins}

\usepackage{ulem}

\usepackage{multirow}
\usepackage{array}

\newcolumntype{x}[1]{>{\centering\hspace{0pt}}p{#1}}

\newcommand{\avg}[1]{\left\langle #1\right\rangle}

\newcommand{\T}[1]{\textrm{#1}}

\begin{document}
 \title{Talent and experience shape competitive social hierarchies}

\author{M\'arton P\'osfai}
\email{posfai@ucdavis.edu}
\affiliation{Complexity Sciences Center and Department of Computer Science, University of California, Davis, CA 95616, USA}
\author{Raissa M. D'Souza}
\affiliation{Complexity Sciences Center, Department of Computer Science and Department of Mechanical and Aerospace Engineering, University of California, Davis, CA 95616, USA}
\affiliation{{Santa Fe Institute, 1399 Hyde Park Road, Santa Fe, NM 87501,  USA}}

\date{\today}

\begin{abstract}
Hierarchy of social organization is a ubiquitous property of animal and human groups, linked to resource allocation, collective decisions, individual health, and even to social instability. Experimental evidence shows that both intrinsic abilities of individuals and social reinforcement processes impact hierarchies; existing mathematical models, however, focus on the latter. Here, we develop a rigorous model that incorporates both features and explore their synergistic effect on stability and the structure of hierarchy. For pairwise interactions, we show that there is a trade-off between relationship stability and having the most talented individuals in the highest ranks. Extending this to open societies, where individuals enter and leave the population, we show that important societal effects arise from the interaction between talent and social processes: (i)~despite positive global correlation between talent and rank, paradoxically, local correlation is negative, and (ii)~the removal of an individual can induce a series of rank reversals. We show that the mechanism underlying the latter is the removal of an older individual of limited talent, who nonetheless was able to suppress the rise of younger, more talented individuals. 
\end{abstract}

\maketitle


\paragraph{Introduction.}\label{sec:introduction}

Hierarchy is a central organizing principle of complex systems, manifesting itself in various forms in biological, social, and technological systems~\cite{Zafeiris2018}. Therefore to understand complex systems, it is crucial to quantitatively describe hierarchies~\cite{Trusina2004,Fushing2011, Mones2012,DeBacco2017} and to identify their origins and benefits~\cite{Bonabeau1995,Nepusz2013}. Among the various forms of hierarchy, here we are concerned with social hierarchies emerging through competition, including dominance and status hierarchies or socioeconomic stratification~\cite{Gould2002,Hsu2005}. Ultimately, such hierarchy represents a ranking of individuals based on social consensus: a high ranking individual is expected to win a conflict against a low ranking one. This type of organization is present in societies ranging from insects to primates and humans~\cite{Chase1980, Fushing2011, Mesterton-Gibbons2016,Lerner2017}, and has been linked to resource allocation, individual  health, collective decisions, and social stability~\cite{Nepusz2013,Tung2012,Beisner2015a,Vandeleest2016}.

The prevalence of social hierarchies motivated a long history of theoretical research in statistical physics and mathematical biology~\cite{Landau1951,Chase1974,Bonabeau1995,Ben-Naim2005,Castellano2009}. The unifying theme in explaining the emergence of hierarchies is positive reinforcement of differences known as the winner effect: initially equally ranked individuals repeatedly participate in pairwise competitions, and after an individual wins, the probability that they win later competitions increases. Conditions for hierarchies to emerge under this mechanism and their structure has been thoroughly investigated~\cite{Ben-Naim2005, Ben-Naim2006,Castellano2009, Mesterton-Gibbons2016, Hsu2005}.

Yet, from experiments focusing on animal groups, we known that in addition to social reinforcement, intrinsic attributes also play a critical role in hierarchy formation~\cite{Hsu2005,Mesterton-Gibbons2016}. The relative strength of the two effects depends on context; however, it was observed that they both affect hierarchies ranging from species with relatively simple social interactions, such as cichlid fish~\cite{Chase2002}, to species that form highly complex societies, such as primates~\cite{Tung2012,Seil2017}.

Despite the clear indication from experiments that both talent and reinforcement matter, we are lacking general theoretical understanding of their synergistic impact~\cite{Beacham2003,Blumm2012}. Here, we develop a rigorous model incorporating both and show that this captures a much richer landscape. For pairwise interactions, we show a trade-off between relationship stability and having more talented individuals be the high-ranked leaders. We then extend the model to open populations, where individuals enter and leave the group, and we characterize both the global and the local structure of hierarchies.

Another pressing issue is to understand the response of hierarchical structure to perturbation, e.g., the effect of removing an individual. In particular, animal behavior experts must often make strategic decisions to remove individuals from captive societies due to health issues or in attempt to promote social stability, which sometimes lead to unanticipated reorganization of hierarchy and even societal collapse~\cite{McCowan2008,Beisner2015a}. We show herein that if either talent or social reinforcement dominate hierarchy formation, the associated models predict smooth response and no rearrangement. It is only if their effects are equally important, that removal of an individual can lead to a non-trivial series of rank reversals.

\paragraph{Model.}
\label{sec:model}

Our starting point is a classic model by Bonabeau {\it et al.} that considers only social reinforcement~\cite{Bonabeau1995}. It describes a group with $N$ members, where the rank of each member is determined by its ability to defeat others in pairwise competitions. This ability is quantified by a score $x_i(t)$, where the subscript indexes the individuals. The scores are initially identical ($x_i(t=0)\equiv 0$) and they change through two discrete-time processes. First, through positive feedback: In each time step, participants are randomly paired to compete with each other, and the winner increases its score by $\delta$. Individual $i$ wins against $j$ with probability
\begin{equation}\label{eq:original-winningprob}
Q_{ij}(t)=\frac{1}{1+\exp[-\beta(x_i(t)-x_j(t))]},
\end{equation}
where $\beta$ is an inverse temperature-like parameter: for large $\beta$ the outcome of the fight is deterministic, for $\beta=0$ both parties have equal chance to win. The second process is forgetting: The effect of a fight wears off exponentially, i.e., $x_i(t)$ is reduced by $\mu x_i(t)$ ($0\leq\mu\leq 1$) in each time step. Describing the full process with the deterministic equation
\begin{equation}\label{eq:original_dynamics}
x_i(t+1) = (1-\mu)x_i(t)+\frac{\delta}{N-1}\sum_{j\neq i}Q_{ij}(t),
\end{equation}
it was shown that, depending on the relative strength of reinforcement and decay, the model supports either  egalitarian ($x_i\equiv 0$) or hierarchical ($x_i\not\equiv 0$) steady state solutions~\cite{Bonabeau1995,Lacasa2006}.

To introduce intrinsic attributes, we offset the score of each participant in Eq.~(\ref{eq:original-winningprob}) by base abilities $b_i$ and $b_j$:
 \begin{equation}\label{eq:intrinsic1}
Q_{ij}(t)=\frac{1}{1+\exp[-\beta(x_i(t)+b_i-x_j(t)-b_j)]}.
\end{equation}
Parameter $b$ quantifies talents that are independent of social processes, yet are relevant to conflict outcomes, such as strength or intelligence. This modification, although formally simple, requires new mathematical description and leads to series of non-trivial behaviors and unanticipated emergent properties.

\begin{figure}[t!]
	\centering
	\includegraphics[width=0.45\textwidth]{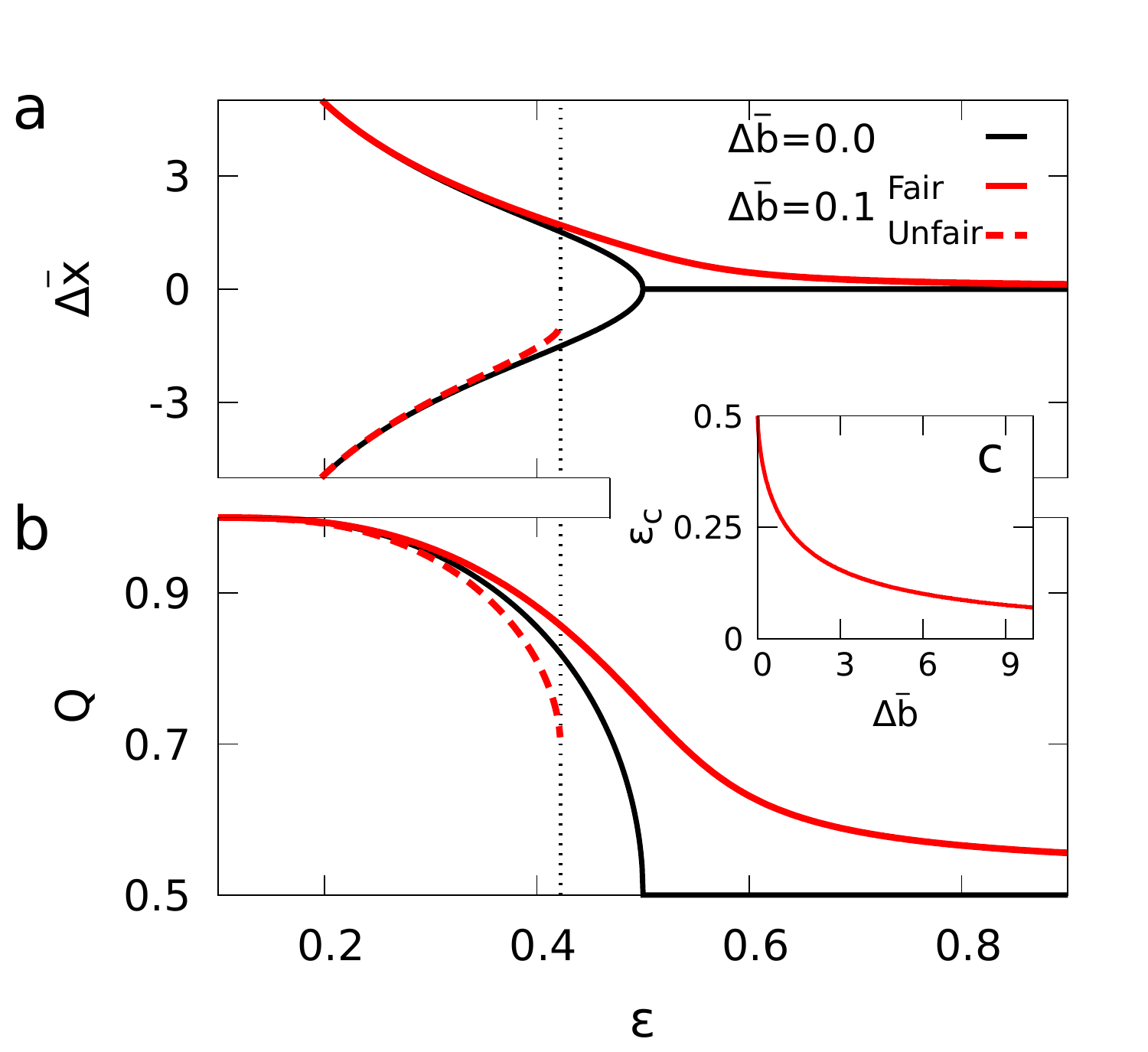}
	\caption{{ \bf Fairness and stability ($N=2$).} {\bf(a)}~Score difference as a function of $\epsilon$, without (black) and with difference in talent (red). If $\Delta b > 0$, for large $\epsilon$ only one hierarchical solution exists corresponding to the fair ranking, i.e., rank is determined by talent (solid); and through a discontinuous transition at $\epsilon_\T c$ (vertical line) a new solution emerges corresponding to the opposite, unfair ordering (dashed). {\bf(b)}~The probability $Q$ that the dominant defeats the subordinate quantifies the stability of a hierarchical relationship; as $\epsilon$ decreases, social stability increases. Shown for the fair (solid) and unfair (dashed) states. {\bf(c)}~Critical point, $\epsilon_\T c$ as a function of $\Delta\bar{ b}$.}
	\label{fig:N2}
\end{figure}

\paragraph{Two individuals.} To understand the consequences of intrinsic differences, it is insightful to first investigate a population of $N=2$. The deterministic equation describing the steady state is
\begin{equation}\label{eq:ss}
0 = -\mu\Delta x + \delta \left( \frac{2}{1+\exp\left[-\beta(\Delta x +\Delta b)\right]}-1\right),
\end{equation}
where $\Delta x = x_1-x_2$ and $\Delta b=b_1-b_2\geq0$. Introducing dimensionless quantities $\Delta \bar{x}=\beta\Delta x$, $\Delta \bar{b}=\beta\Delta b$ and $\epsilon=\mu /(\delta \beta)$ leads to 
\begin{equation}\label{eq:singlepar}
0 = -\epsilon \Delta\bar{x} + \frac{2}{1+\exp\left[-\Delta \bar{x} -\Delta \bar{b}\right]}-1,
\end{equation}
meaning that the steady state is determined by the talent difference and a single parameter $\epsilon$ measuring the relative strength of decay to social reinforcement~\footnote{It is interesting to note that introducing $m=\mu\Delta x$ and  $\tilde \beta =\beta\delta/(2\mu)$ maps Eq.~(\ref{eq:ss}) to the meanfield Ising model and the intrinsic difference maps to a non-zero external field $h=\mu\Delta b/\delta$.}.

Systematically changing $\epsilon$, we observe a transition at $\epsilon_\T c(\Delta\bar{ b})$ separating regimes with one and two stable solutions; the nature of the transition depends on the presence of intrinsic differences. If $\Delta\bar{ b}=0$ (Fig.~\ref{fig:N2}a black line), we recover the original model: For $\epsilon>\epsilon_\T c (0)$ we find one solution, representing the egalitarian state $\Delta\bar{ x}= 0$, and at $\epsilon_\T c(0)$ two symmetric hierarchical solutions ($\Delta\bar{ x}_1=-\Delta\bar{ x}_2\neq 0$) emerge through a pitchfork bifurcation. If $\Delta\bar{ b}> 0$ (Fig.~\ref{fig:N2}a red line): For $\epsilon>\epsilon_\T c(\Delta\bar{ b})$ we again find just one solution; this solution, however, is not egalitarian ($\Delta\bar{ x}> 0$), but it is ``fair'' in that the more talented individual outranks the less talented. At $\epsilon_\T c(\Delta\bar{b})$ a new stable solution appears through a discontinuous transition supporting the opposite order, which is ``unfair'', meaning that the less talented outrank the more talented. In other words, social reinforcement can outpace intrinsic abilities.  We call the $\Delta\bar{ x}>0$ solution ``fair'' and the $\Delta\bar{ x}<0$ one ``unfair'', since high-ranked individuals tend to have better access to resources, more impact on collective decisions, and higher chance to foster offspring.

Figure~\ref{fig:N2}c shows the dependence of $\epsilon_\T c$ on $\Delta\bar{ b}$. In general, no closed-form solution is available; limiting cases, however, can be worked out analytically: for small differences we find $(\epsilon_\T c-1/2)\sim \Delta\bar{ b}^{2/3}$ and for large differences $\epsilon_\T c = {\Delta\bar{ b}}^{-1}$. The latter indicates that increasing talent difference or decreasing reinforcement pushes the system to a regime where only the fair solution exists. Since the fair solution intuitively benefits society, this prompts the question: what is the role of social reinforcement?

To answer this question, we quantify the stability of a dominant-subordinate relationship with $Q$, the probability that the dominant wins a conflict, $Q\approx 1/2$ indicates an unstable relationship and $Q\approx 1$ a well-defined relationship. Stable relationships reduce overall aggression and are positively associated with individual health~\cite{Vandeleest2016}. Figure~\ref{fig:N2}b shows that strong social reinforcement (high $\delta$ and thus low $\epsilon$) increases $Q$, revealing a fundamental trade-off between stability and fairness: stable relationships require strong social reinforcement; however, strong reinforcement allows for unfair hierarchical states. Similar trade-off was experimentally observed in rankings of products in a marketplace competing for the attention of consumers: strong social reinforcement led to less accuracy in selecting the highest quality product, and to larger differences in market share~\cite{Salganik2006}.

\paragraph{Open populations.}
\label{sec:large_pop}
So far we focused on the relationship of two individuals, now we turn our attention to larger, changing populations. We study groups of $N$ individuals where the talent of each individual is drawn randomly from a distribution $p(b)$. We initially allow the population to reach a stable ranking. Then in each step, we remove a random individual and add a new member $i$ to the bottom of the society, i.e., $x_i=0$, and again allow the population to reach a stable ranking.

For simplicity we restrict our investigation to the $\beta\rightarrow\infty$ limit, in which case $Q_{ij}$ becomes a step function. This allows us to explicitly formulate the condition for two consecutively ordered individuals to reverse ranks during the evolution of the hierarchy~\cite{SM}:
\begin{equation}\label{eq:cond_to_pass}
b(k+1) - b(k) > \Delta x\equiv \frac{\delta}{\mu(N-1)},
\end{equation}
where $b(k)$ is the talent of the individual ranked $k$th (note that $k=1$ is the top and $k=N$ is the bottom rank); and $\Delta x$ is the score difference of two consecutively ranked individuals $x(k)-x(k+1)$ which turns out to be independent of their ranks~\cite{SM}. Therefore, $\Delta x$ is the additional talent needed to overcome the advantage of having higher rank. Parameters $\delta$ and $\mu$ only effect the system through $\Delta x$; therefore treating $\Delta x$ as a parameter completely specifies the dynamics. The $\beta\rightarrow\infty$ limit allows us to study a simplified representation of the dynamics in Eqs.~(\ref{eq:original_dynamics}) and (\ref{eq:intrinsic1}): We check each consecutively ranked individual and if Eq.~(\ref{eq:cond_to_pass}) is satisfied, we reverse their order; we repeat this until no more pairs are reversed. In the Supplemental Material, we derive various properties of the hierarchy through exact combinatorics and meanfield-like approximations~\cite{SM}.

\begin{figure}[t!]
	\centering
	\includegraphics[width=0.45\textwidth]{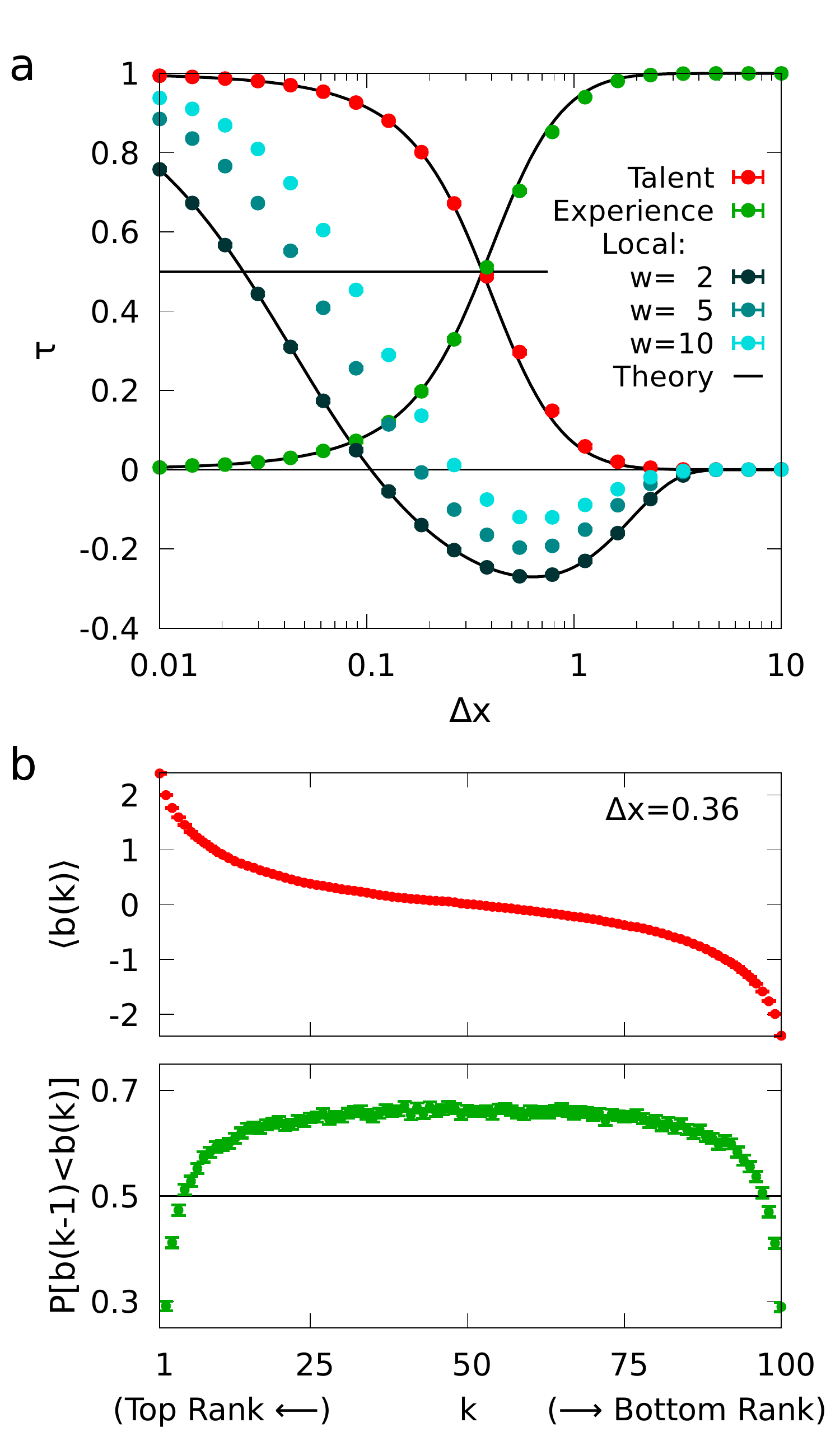}
	\caption{{ \bf Talent-rank correlation.} {\bf(a)}~Kendall's tau as a function of $\Delta x$. Global talent-rank (red) and experience-rank (green) correlation shows a crossover between talent and experience dominated limiting cases. Counter-intuitively, we find that locally talent and rank are anti-correlated (blue) as shown for local windows of increasing size $w$. {\bf(b)}~Local rank-talent anti-correlation. In the crossover regime, the expected talent increases with rank (red), yet the probability that an individual's immediate superior is less talented is greater than 1/2 (green). In (a) and (b), results are shown for populations of $N=100$, continuous lines are analytical solutions~\cite{SM}. Data points are simulations of the dynamics defined in Eq.~(\ref{eq:cond_to_pass}), representing an average of 10,000 independent samples and error bars provide the 95\% CI.}
	\label{fig:anticorr}
\end{figure}

The talent $b$ of an individual represents an intrinsic ability or a combination of abilities that influence the outcome of a fight. In our analysis we derive a number of properties of social hierarchies for general continuous talent distribution $p(b)$, including heavy-tailed distributions. Whenever specific $p(b)$ is necessary for calculations or simulations, we focus on the standard normal distribution. Indeed, body size, intelligence, and other relevant abilities are often normally distributed. 

We now systematically investigate the structure of the emergent hierarchy as a function of $\Delta x$, the additional talent difference needed to overcome rank difference. We measure correlation between rank and talent ($\tau_\T{tal}$) and between rank and experience ($\tau_\T{exp}$) using Kendall's tau coefficient, where experience is the amount of time an individual has spent in the population. For example, $\tau_\T{tal}=1$ indicates talent completely determines rank and $\tau_\T{tal}=0$ indicates no correlation. Analytical calculations and simulations show that for large $\Delta x$, rank is dominated by experience, meaning that the only way to advance in the hierarchy is if a higher ranking individual is removed; and for small $\Delta x$ rank is dominated by talent (Fig.~\ref{fig:anticorr}a). These two limiting cases are separated by a regime where both talent and experience matter, theory predicts that the crossover point, where $\tau_\T{tal}=\tau_\T{exp}=1/2$, is $\Delta x_\T c\approx 0.36$ for $N=100$.

Experimental measurement of $\tau_\T{tal}$ is challenging since it requires exact identification of the relevant talents; determining $\tau_\T{exp}$, however, is straight forward. Indeed, Tung {\it et al.} established small captive groups of macaques by introducing animals one-by-one into an enclosure and found that the Spearman's correlation between rank and experience is $\rho_\T{exp}=0.61$, demonstrating that some real systems are in fact near the crossover point~\cite{Tung2012}.

In addition to global correlations, we also quantify local orderedness by calculating $\tau_\T{tal}(w)$, the talent-rank correlation averaged over a sliding window of length $w$. Counter-intuitively, Fig.~\ref{fig:anticorr}a shows that in the crossover regime $\tau_\T{tal}(w)$ is negative, meaning that locally rank and talent are anti-correlated. Figure~\ref{fig:anticorr}b provides an additional aspect of this paradox situation: The expected talent $\avg{b(k)}$ of an individual ranked $k$th at a random time step monotonically increases with rank; yet the probability that the $(k-1)$th individual, the one immediately outranking the $k$th, is less talented than the $k$th is greater than $1/2$.

To understand the mechanism producing the local anti-correlation, first consider two consecutive individuals forming an ordered pair with respect to talent, i.e., $b(k)<b(k-1)$. If a new individual arrives with talent $b$ such that $b(k)+\Delta x < b$ and $b(k-1)<b<b(k-1)+\Delta x$, it can pass the $k$th individual, but cannot pass the $(k-1)$th, lodging itself between the two and creating an unordered pair. Once an unordered pair exists, i.e., $b(k)>b(k-1)$, any individual passing the $k$th will necessarily pass the $(k-1)$th too. Therefore an unordered pair will remain unordered until one of the pair is removed. This asymmetry in creating ordered and unordered pairs is responsible for the local anti-correlation.

\begin{figure}[t!]
	\centering
	\includegraphics[width=0.45\textwidth]{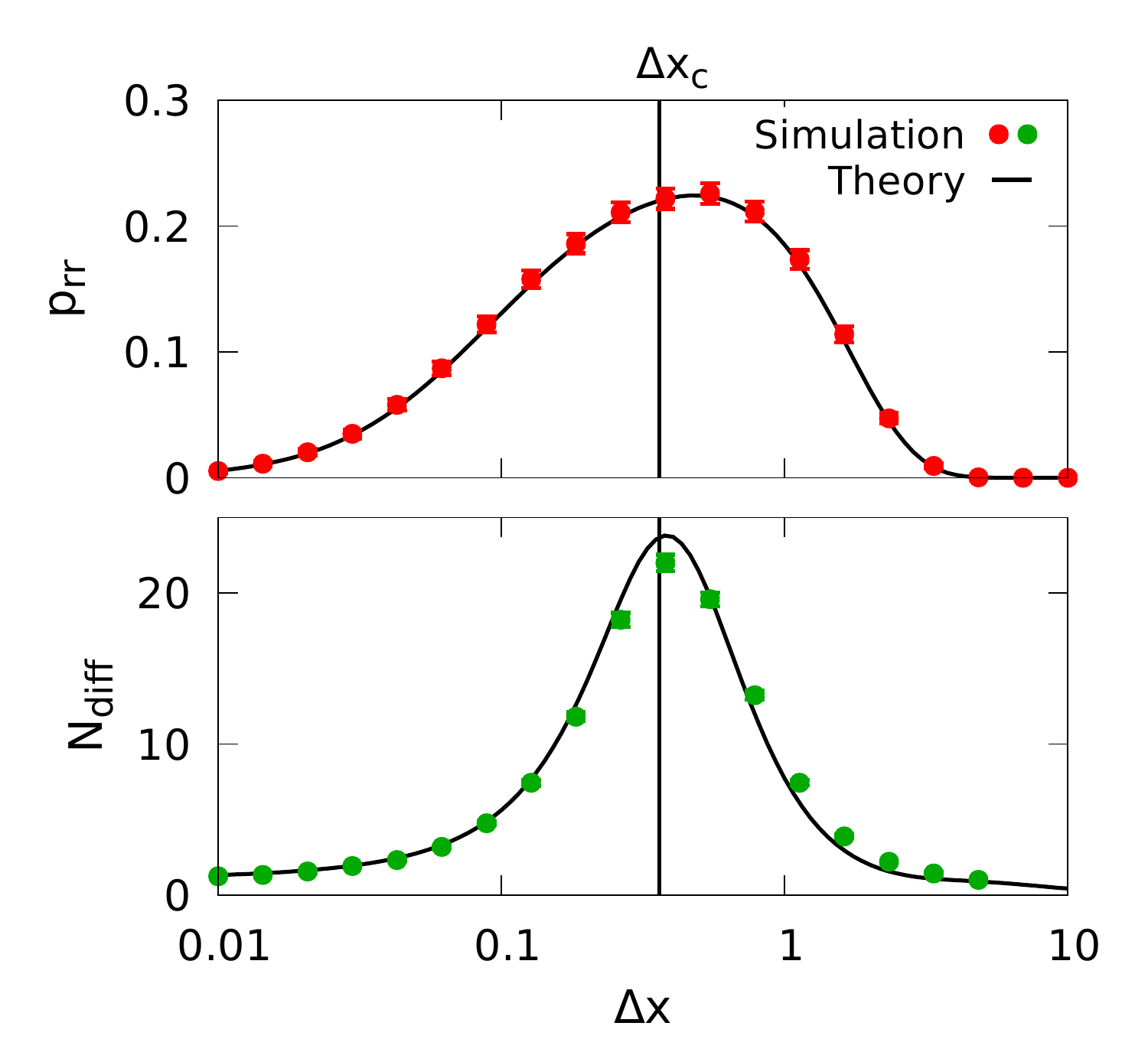}
	\caption{{ \bf Removal of a group member.} In the limiting cases of talent and experience dominated societies, removal has trivial effect, while in the crossover regime the removal of an individual causes rank reversals with finite probability (red). The average number of rank reversals $N_\T{diff}$ peaks near $\Delta x_\T{c}$ (green). Results are shown for populations of $N=100$, continuous lines are analytical solutions~\cite{SM}. Data points are are simulations of the dynamics defined in Eq.~(\ref{eq:cond_to_pass}), representing an average of 50,000 time steps and error bars provide the 95\% CI.}
	\label{fig:removal}
\end{figure}

Finally, we also investigate the effect of removing an individual. We find that in the talent or experience dominated limiting cases the system's response is trivial and no re-organization happens. However, Figure~\ref{fig:removal} shows that $p_\T{rr}$, the probability that removal of an individual induces rank reversals, is non-zero in the crossover regime. For $N=100$, both $p_\T{rr}$ and the average number of these rank reversals $N_\T{diff}$ peak near, but not exactly at, the crossover point $\Delta x_\T{c}$. For removal-induced rank reversals to happen, at least three consecutively ranked individuals are needed in opposite order with respect to talent, i.e., $b(k+1) > b(k) > b(k-1)$. If the condition $b(k+1) - b(k-1)>\Delta x$ is satisfied, the removal of the $k$th individual allows the $(k+1)$th to pass the $(k-1)$th, which can lead to a series of rank reversals. In other words, the $k$th individual is not talented enough to further advance in society, but is capable of holding back a younger, more talented contender.

Understanding the response of hierarchies to external perturbation is an important issue. Particularly, removal of animals from primate groups can sometimes lead to large shifts in hierarchy and instabilities endangering the group~\cite{McCowan2008,Beisner2015a}. Here we demonstrated that traditional models of hierarchy formation, those only considering either intrinsic abilities or social feedback, predict trivial response to removal, and that both effects have to be present simultaneously to observe rank reversals.

So far we have focused on populations of $N=100$ individuals. In the Supplementary Material, we extract the scaling behavior of various properties for large $N$~\cite{SM}. Local quantities, such as $\tau_\T{tal}(2)$ and $p_\T{rr}$, scale as $\tau_\T{tal}(2)=\tau^{(1)}_\T{tal}(2, N\Delta x)$ and $p_\T{rr}=p^{(1)}_\T{rr}(N\Delta x)$ for small $\Delta x$; and are independent of $N$ for large $\Delta x$, i.e., $\tau_\T{tal}(2)=\tau^{(2)}_\T{tal}(2, \Delta x)$ and $p_\T{rr}=p^{(2)}_\T{rr}(\Delta x)$. The location of their extreme value is at the crossover of these two regimes and in case of normal talent distribution scales as $\sim (\ln N)^{1/6}/N^{1/3}$. We find universal bounds
\begin{equation}
\begin{split}
\tau_\T{tal}(2)&\geq-2\ln2+1,\\
p_\T{rr}&\leq 0.294\ldots,
\end{split}
\end{equation} 
for any continuous unbounded talent distribution, and these bounds are reached in the large population limit. For global talent correlation, on the other hand, we find that $\tau_\T{tal}\rightarrow 1$ if $\sqrt{N}\Delta x\rightarrow 0$ and $\tau_\T{tal}\rightarrow 0$ if $\sqrt{N}\Delta x\rightarrow \infty$. Therefore, the crossover point where $\tau_\T{tal}=\tau_\T{exp}=1/2$ scales as $\Delta x_\T c\sim 1/\sqrt{N}$. The average number of rank reversals $N_\T{diff}$ depends on global correlations, and peaks near the crossover point $\Delta x_\T c$. Note that in the parametrization of the model, provided in Eq.~(\ref{eq:cond_to_pass}), $\Delta x=\delta/[\mu(N-1)]$; meaning that for $N\rightarrow\infty$, global correlation becomes talent dominated and local correlation may become negative depending on the value of $N\Delta x$. Other scalings of $\Delta x$ are also possible through adjustment of $\delta$ or $\mu$, or if individuals do not randomly select opponents, but selectively compete with similarly ranked ones. The properties we observed for finite hierarchies may become more pronounced in the large population limit, for example, if $N\Delta x \rightarrow \infty$ but $\sqrt{N}\Delta x \rightarrow 0$,  global correlation $\tau_\T{tal}$ converges to one, while local correlation approaches its theoretical minimum. In Table~\ref{tab:limits}, we provide detailed enumeration of possible behavior in the large population limit assuming $\Delta xN^\alpha=C$, where $C>0$ is constant.

\renewcommand{\arraystretch}{1.}
\begin{table}[t]
\centering
\begin{tabular}{c|x{2.2cm}x{2.2cm}x{2.2cm}x{2.2cm}x{2.2cm}}
& \multicolumn{5}{ c }{In the limit of $N\rightarrow\infty$} \tabularnewline[-2ex]
& $\tau_\T{tal}$ & $\tau_\T{exp}$ &  $\tau_\T{tal}(2)$ & $p_\T{rr}$ & $N_\T{diff}/N$ \tabularnewline
\hline\hline
$1<\alpha$ & 1 & 0 & 1 & 0 & 0 \tabularnewline
\hline
$\alpha=1$ & 1 & 0 & $\tau^{(1)}_\T{tal}(2,C)$ & $p^{(1)}_\T{rr}(C)$ & 0 \tabularnewline
\hline
$1/2<\alpha<1$ & 1 &  0 & $-2\ln 2+1$ & $0.294\ldots$ & 0 \tabularnewline
\hline
$\alpha=1/2$ & $\tau_\T{tal}(C)$ & $1-\tau_\T{tal}(C)$ & $-2\ln 2+1$ & $0.294\ldots$ & $f(C)$ \tabularnewline
\hline
$0<\alpha<1/2$ & 0 & 1 & $-2\ln 2+1$ & $0.294\ldots$ & 0 \tabularnewline
\hline
$0=\alpha$ & 0 & 1 & $\tau^{(2)}_\T{tal}(2,C)$ & $p^{(2)}_\T{rr}(C)$ & 0 \tabularnewline
\hline
$\alpha<0$  & 0 & 1 & 0 & 0 & 0
\end{tabular}
\caption{\label{tab:limits} {\bf Structure of hierarchy in the large population limit assuming $\Delta x N^\alpha= C$.} The numerical values are valid for any continuous unbounded talent distribution, while the scaling functions are specific to the talent distribution and are calculated in the Supplementary Material~\cite{SM}. }
\end{table}

\paragraph{Discussion.}

We studied the synergistic effect of talent and social reinforcement on the structure of competitive social hierarchies, and we identified behaviors that cannot be observed if either effect dominates. Although we derived our model assuming pairwise conflicts and a winner effect, we believe that the results can be interpreted more generally: (i)~The mechanism behind both local talent-rank anti-correlation and removal-induced rank reversals is that to pass someone in rank it is not enough to be more talented, but the talent difference has to be sufficient to compensate for the advantage of being higher ranked -- a process relevant to many systems, examples might include rankings of scientists, best seller lists, or sports rankings. (ii)~We introduced parameter $b$ to capture individual talents; however, it can be thought of as a proxy for support of kin or as a simplified model of reputation received in exchange for non-adversarial social interactions.

Finally, our results prompt many research questions, both experimental and theoretical. For example, local anti-correlation and removal-induced rank reversals are predictions that are testable through experiments. Future theoretical work may investigate sources of complexity not captured by our model, for example, the role of aging or slow deterioration of talent, or non-linear hierarchies, where social tiers are occupied by multiple individuals.

\begin{acknowledgments}
\paragraph{Acknowledgements.} We thank Brenda McCowan, Brianne Beisner, Darcy Hannibal, and Kelly Finn for useful discussions. We gratefully acknowledge support from the US Army Research Office MURI Award No. W911NF-13-1-0340
and the DARPA Award No. W911NF-17-1-0077.
\end{acknowledgments}

\bibliographystyle{apsrev4-1}

%

\end{document}


\title{Supplementary Material -- Talent and experience shape competitive social hierarchies}

\author{M\'arton P\'osfai}
\email{posfai@ucdavis.edu}
\affiliation{Complexity Science Center and Department of Computer Science, University of California, Davis, CA 95616, USA}
\author{Raissa M. D'Souza}
\affiliation{Complexity Science Center, Department of Computer Science and Department of Mechanical and Aerospace Engineering, University of California, Davis, CA 95616, USA}
\affiliation{{Santa Fe Institute, 1399 Hyde Park Road, Santa Fe, NM 87501,  USA}}

\date{\today}

\maketitle

\setcounter{tocdepth}{0}
\tableofcontents

\section{Introduction}

In the Supplementary Material, we provide in detail the analytical solution of the model of hierarchy formation in large, dynamically changing populations.
In Sec.~\ref{sec:dynamics}, we introduce a simplified description of the dynamics of the model in the $\beta\rightarrow\infty$ limit and our overall approach that ultimately allows analytical solution. 
In the following four sections, we derive and analyze the global rank-talent ($\tau_\T{tal}$) and rank-experience ($\tau_\T{exp}$) correlation, the local rank-talent correlation for window size $w=2$ ($\tau_\T{tal} (2)$), the probability of removal-induced rank reversals ($p_\T{rr}$), and the expected number of pairwise rank reversals ($N_\T{diff}$). We identify the scaling behavior of each quantity for large population sizes. Finally, in Sec.~\ref{sec:Ninfty}, we enumerate the possible emergent hierarchies and identify their properties in the infinite population limit. 

\section{Dynamics in the $\beta\rightarrow\infty$ limit and analytical approach}\label{sec:dynamics}

In this section, we setup the framework that allows us to analytically solve the model for $N>2$ populations where individuals may leave and enter the hierarchy.

Consider a group of $N$ individuals indexed $i=1,2,3,\dots,N$, each individual is characterized by intrinsic ability or talent $b_i$ drawn from a given random distribution $p(b)$. We initially allow the population to reach a stable ranking. Then in each step, we remove a random individual and add a new member $j$ with talent $b_j$ to the bottom of the society, i.e., $x_j=0$, and again allow the population to reach a stable ranking. 

For simplicity we restrict our investigation to the $\beta\rightarrow\infty$ limiting case, in which case $Q_{ij}$ becomes a step function, meaning that $i$ always wins if $x_i+b_i>x_j+b_j$. In a stable ranking, the score of the $k$th individual, denoted $x(k)$, is simply
\begin{equation}
x(k)= (N-k)\frac{\delta}{\mu(N-1)}.
\end{equation}
Note that confusingly ``high rank'' corresponds to small values of $k$, e.g., $k$ is higher ranked than $k+1$. 

The condition for two consecutively ranked individuals to reverse ranks during the temporal evolution of the hierarchy is
\begin{equation}\label{eq:cond_to_pass_A}
\begin{split}
b(k+1)+x(k+1)-b(k)-x(k)&>0,\\
b(k+1) - b(k) > x(k)-x(k+1)&\equiv\Delta x=\frac{\delta}{\mu(N-1)},
\end{split}
\end{equation}
where $b(k)$ is the intrinsic ability of the individual ranked $k$th and $\Delta x$ is the score separation between two consecutive individuals. This $\Delta x$ is independent of $k$, and parameters $\delta$ and $\mu$ only affect the stable ranking through $\Delta x$; therefore, we treat it as a parameter of the dynamics. The state of the system is completely described by the ordering of individuals $O=[o(1),o(2),\ldots,o(N)]$, where $o(k)$ is the index of the individual ranked $k$th. 

After adding a new individual to the bottom of the hierarchy, we allow the system to reach a stable ranking, i.e., we allow the newcomer to rise as its talent allows. As briefly described in the main text, instead of using the dynamics defined in Eqs.~(2) and (3) of the main text to reach the stable ranking, the condition in Eq.~(\ref{eq:cond_to_pass_A}) allows us to employ a simplified representation of the dynamics: We check each consecutive pair $[o(k),o(k+1)]$ ($k=1,2,\ldots,N-1$), and if condition Eq.~(\ref{eq:cond_to_pass_A}) is satisfied we reverse their order. We repeat this until no further change is found. All results for $N>0$ populations, reported  here and in the main text rely on this simplified representation. 

One further observation is required to analytically solve the model: the probability of observing the ordering $O$ is equal to the probability of creating the same ordering by starting from a single individual and adding new members one-by-one. More precisely: At time step $t=1$, we start with one individual with intrinsic ability $b_1$. At the time step $t=2$, we add one new individual with intrinsic ability $b_2$ to the bottom of the society. We repeat this step until the population reaches $N$ individuals. In the following, we calculate properties of rankings obtained from this alternative process -- due to the equivalence these properties will hold for the original model as well.

Note that here we use $b(k)$ to denote the talent of an individual ranked $k$th and $b_t$ to denote the talent of an individual added at time step $t$.

\section{Global rank-talent ($\tau_\T{\normalfont tal}$) and rank-experience ($\tau_\T{\normalfont exp}$) correlation}\label{sec:global_corr}

In this section, we calculate the correlation between rank from the dynamical process and ranking based on talent or experience. We measure the correlation using the Kendall's tau coefficient, which is defined as
\begin{equation}\label{eq:tau}
\tau = \frac{n_+ - n_-}{N(N-1)/2},
\end{equation}
where $n_+$ counts the number of pairs that are in the same order in the two rankings that we compare, and $n_-$ counts the number of pairs that are in the opposite order. If the two rankings are exactly the same, $\tau=1$; if there is no correlation, $\tau=0$; and if the two rankings are exact opposites, $\tau=-1$.

We introduce $N_ -^\T{tal}(t)$ as the number of individuals that arrived after $t$, are more talented than $b_t$, but ultimately receive a lower rank. Therefore we can express the number of discordant pairs with respect to rank and talent as $n_-=\sum_{t=1}^N N_-^\T{tal}(t)$. Therefore Eq.~(\ref{eq:tau}) for rank-talent correlation becomes
\begin{equation}\label{eq:tau_b}
\tau_\T{tal} = 1-4\sum_{t=1}^N\frac{  N_-^\T{tal}(t)}{N(N-1)}.
\end{equation}

To calculate $N_ -^\T{tal}(t)$, we first define the threshold $a(t^\prime-t,b_t)$, which gives the talent value below which an individual arriving at $t^\prime>t$ cannot pass the individual that arrived at $t$ with talent $b_t$. According to Eq.~(\ref{eq:cond_to_pass_A}), initially talent at least $a(1,b_t) = b_t+\Delta x$ is needed to pass. This threshold increases to $b_{t+1}+\Delta x$, if a new individual arrives with talent $b_{t+1}$, such that $b_{t}<b_{t+1}<b_{t}+\Delta x$; otherwise it remains unchanged. To approximate the expectation value of $a(t^\prime-t,b_t)$, we can write the following recursion
\begin{equation}\label{eq:a}
\begin{split}
a&(t^\prime+1-t,b_t) = a(t^\prime-t,b_t) +\!\!\mkern-18mu \int\limits_{a(t^\prime-t,b_t)-\Delta x}^{a(t^\prime-t,b_t)}\mkern-18mu\!\! p(b)(b+\Delta x-a(t^\prime-t,b_t))db,\\
a&(t+1,b_t) = b_t +\Delta x. 
\end{split}
\end{equation}
We can now calculate
\begin{equation}\label{eq:N-tal}
N_-^\T{tal}(t)= \int_{ } db_t p(b_t)\sum_{t^\prime=t+1}^{N}\!\!\!\!\int\limits_{b_{t}}^{a(t^\prime-t,b_t)}\!\!\!\!\!\!p(b)db,
\end{equation}
where the summand is the probability that the individual arriving at $t^\prime$ is more talented than $b_t$ but cannot pass it. We can now calculate $\tau_\T{tal}$ by plugging in $N_-^\T{tal}(t)$ into Eq.~(\ref{eq:tau_b}).

We calculate $\tau_\T{exp}$ in a similar fashion. By defining $N_-^\T{exp}(t)$ as the number of individuals that arrived after $t$ and passed the individual that arrived at $t$; Eq.~(\ref{eq:tau}) for rank-experience correlation becomes
\begin{equation}\label{eq:tau_exp}
\tau_\T{exp} = 1-4\sum_{t=1}^N\frac{  N_-^\T{exp}(t)}{N(N-1)},
\end{equation}
and we can write
\begin{equation}
N_-^\T{exp}(t)= \int_{ } db_t p(b_t)\sum_{t^\prime=t+1}^{N}\int\limits_{a(t^\prime-t,b_t)}^{\infty}\!\!\!\!\!\!p(b)db,
\end{equation}
together with Eq.~(\ref{eq:tau_exp}) this provides $\tau_\T{exp}$.

It is worth noting that the sum of discordant rank-talent and rank-experience pairs can be evaluated as
\begin{equation}
\sum_{t=1}^N [N_-^\T{tal}(t)+N_-^\T{exp}(t)]= \sum_{t=1}^{N}\sum_{t^\prime=t+1}^{N}\int_{ } db_t p(b_t)\int\limits_{b_t}^{\infty}\!\!p(b)db = \frac{N(N-1)}{4},
\end{equation}
which is independent of $p(b)$. Inserting this into Eq.~(\ref{eq:tau_exp}) we get
\begin{equation}\label{eq:tau_tal-exp}
\begin{gathered}
\tau_\T{exp} = 1-4\frac{ N(N-1)/4 - \sum_{t=1}^NN_-^\T{tal}(t)}{N(N-1)},\\
\tau_\T{exp}=1-\tau_\T{tal}.
\end{gathered}
\end{equation}
Therefore, the coefficients measuring the global talent-rank and talent-experience correlation always sum up to one.

\begin{figure}[t]
	\centering
	\includegraphics[width=.66\textwidth]{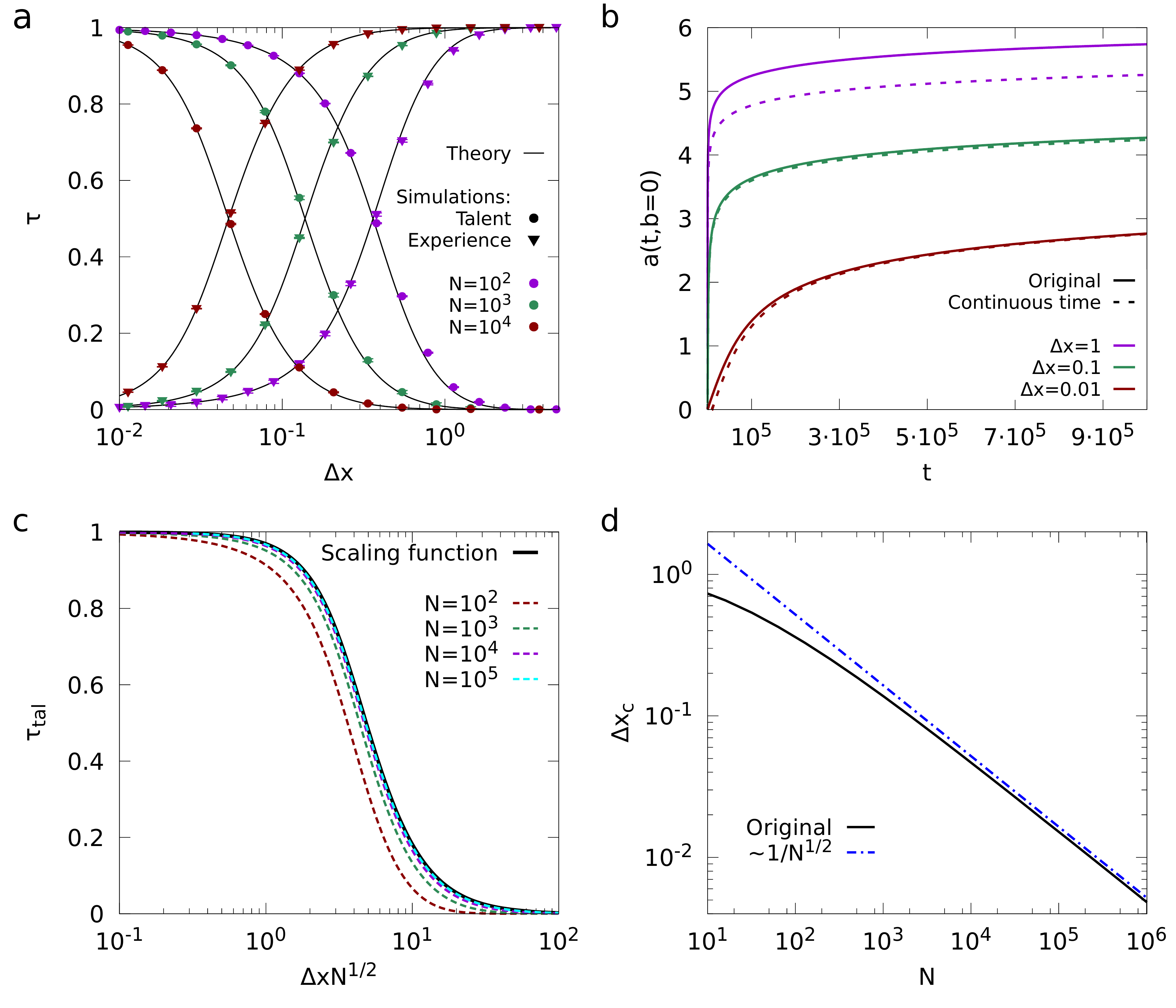}
	\caption{{ \bf Global correlations $\tau_\T{\normalfont tal}$ and $\tau_\T{\normalfont exp}$.} {\bf(a)}~Comparing the analytical solution of $\tau_\T{tal}$ and $\tau_\T{exp}$ (Eqs.~(\ref{eq:a}) and (\ref{eq:N-tal})) to numerical simulations shows excellent agreement for various system sizes. Error bars represent the standard error of the mean. {\bf(b)}~Comparing the solution of $a(t, b=0)$ obtained from solving the original discrete time Eq.~(\ref{eq:a}) and the continuous time approximation Eq~(\ref{eq:a-ct}). The approximation is accurate for large $t$ and small $\Delta x$.  {\bf(c)}~Scaling of $\tau_\T{tal}$. The continuous time approximation predicts that plotting $\tau_\T{tal}$ as a function of $\Delta x \sqrt{N}$ collapses the values of $\tau_\T{tal}$. We compare the rescaled discrete time solution of $\tau_\T{tal}$ to the scaling function is provided by Eq.~(\ref{eq:tau_b_ct}), we find that the scaling is accurate for large populations. {\bf(d)}~The crossover point $\Delta x_\T{c}$, where $\tau_\T{tal}=\tau_\T{exp}=1/2$, scales as $\sim 1/\sqrt{N}$ for large populations.}
	\label{fig:SI-tau}
\end{figure}

We numerically evaluate Eqs.~(\ref{eq:a}) and (\ref{eq:N-tal}) and compare their predictions to simulations. Figure~\ref{fig:SI-tau}a shows excellent agreement for various population sizes. In the following we extract the scaling behavior of $\tau_\T{tal}$, $\tau_\T{exp}$, and the crossover point $\Delta x_\T{c}$ (where $\tau_\T{tal}=\tau_\T{exp}=1/2$) for large population sizes $N$.

\subsection*{Continuous time approximation.}
 
 To extract the scaling behavior of $\tau_\T{tal}$, we approximate Eq.~(\ref{eq:a}) for small $\Delta x$ as
\begin{equation}\label{eq:a-smallDx}
\begin{split}
 a(t+1,b) &= a(t,b)  + \frac{1}{2}p(a(t,b))\Delta x^2,\\
 a(1,b) &= b.
\end{split}
\end{equation}
We can further simplify this using continuous time approximation, leading to 
\begin{equation}\label{eq:a-ct}
\begin{split}
 \dot{a}(t,b) &= \frac{1}{2}p(a(t,b))\Delta x^2,
 \end{split}
\end{equation}
which can be solved by separation of varibles
\begin{equation}\label{eq:a-ct-sol}
\begin{split}
\int_{ }\frac{da}{p(a)} &= \frac{1}{2}\Delta x^2 t.
 \end{split}
\end{equation}
For the standard normal distribution, there is no closed form of $a(t,b)$ available; however, we can immediately note that $a(t,b,\Delta x)\equiv a(\Delta x^2 t,b)$ for any talent distribution. Figure~\ref{fig:SI-tau}b compares the solution of $a(t,b)$ using the original Eq.~(\ref{eq:a}) and the approximations from Eqs.~(\ref{eq:a-smallDx}) and (\ref{eq:a-ct}), showing good agreement for small $\Delta x$ and large $t$.

Next, we approximate the sums in Eq.~(\ref{eq:N-tal}) with integrals and simplify notations:
\begin{equation}
\sum\limits_{t=1}^N N_-^\T{tal}(t)= \int_{ } db p(b)\int\limits_{1}^{N}\!\!dt\int\limits_{t+1}^{N}\!\!dt^\prime\left[P(a(\Delta x^2[t^\prime-t],b))-P(b)\right].
\end{equation}
Switching the order of integration and evaluating integrals whenever it is possible, we obtain
\begin{equation}\label{eq:Ntal_ct}
\sum\limits_{t=1}^N N_-^\T{tal}(t)= N^2\int_{0}^{1}\!\!ds \!\!\int_{ } db p(b) (1-s) P(a(s\cdot\Delta x^2N,b))-\frac{N^2}{4}, 
\end{equation}
where we kept only leading order terms of $N$, and made the substitution $(t^\prime-t)/N\rightarrow s$. Inserting the result into Eq.~({\ref{eq:tau_b}), we get
\begin{equation}\label{eq:tau_b_ct}
\tau_\T{tal} = 2-4\int_{0}^{1}\!\!ds \!\!\int_{ } db p(b) (1-s) P(a(s\cdot\Delta x^2N,b)),
\end{equation}
where the approximation is accurate for large populations. We can now see that
\begin{enumerate}[(i)]
\item The global correlation scales as $\tau_\T{tal}(\Delta x, N)\equiv \tau_\T{tal}(\Delta x^2N)$ for large $N$; therefore plotting $\tau_\T{tal}$ as a function of $\Delta x \sqrt{N}$ collapses the values of $\tau_\T{tal}$ for large populations (Fig.~\ref{fig:SI-tau}c).
\item It is a direct consequence of point (i) that the crossover point $\Delta x_c$, where $\tau_\T{tal}=\tau_\T{exp}=1/2$, scales as $\Delta x_\T{c}\sim 1/\sqrt{N}$ (Fig.~\ref{fig:SI-tau}d).
\item If $\Delta x^2 N\rightarrow 0$ then $\tau_\T{tal}\rightarrow 0$, and if $\Delta x^2 N\rightarrow \infty$ then $\tau_\T{tal}\rightarrow 1$.
\item So far in the calculations, we did not specify $p(b)$; therefore the above properties are true for any continuous unbounded talent distribution.
\end{enumerate}
We have shown in Eq.~(\ref{eq:tau_tal-exp}) that $\tau_\T{exp}=1-\tau_\T{tal}$; therefore the scaling behavior of $\tau_\T{exp}$ and $\tau_\T{tal}$ is identical.

\section{Local rank-talent correlation ($\tau_\T{\normalfont tal}(w=2)$)}
\label{sec:poo}

In this section, we derive the exact formula and scaling behavior of  $\tau_\T{tal}(w=2)$, the talent-rank correlation coefficient averaged over a sliding window of $w=2$, i.e., covering two consecutively ranked individuals. The window size $w=2$ allows us to write
\begin{equation}\label{eq:tauw2}
\tau_\T{tal}(w=2) = 1-2p_\T{oo},
\end{equation}
where $p_\T{oo}$ is the probability that a randomly selected consecutive pair is in opposite order with respect to talent, i.e., the individual ranked higher is less talented. For such pairs to exist the condition \begin{equation}\label{eq:wrong_order}
b_1<b_2<b_1+\Delta x
\end{equation}
most hold, where $b_1$ and $b_2$ are consecutively ranked and $b_1$ is ranked higher. (For brevity, we  refer to ``individual with talent $b_i$'' simply as $b_i$.)

To proceed, we first calculate the probability that $b_2$ is introduced $j+1$ time steps after $b_1$:
\begin{enumerate}
\item {\it $b_1$ arrives} with probability $p(b_1)db_1$;
\item {\it $j$ individuals arrive}, in order for $b_2$ to be consecutive with $b_1$, all $j$ individuals have to either pass $b_1$ or be passed by $b_2$,  the probability of this is $\left(1 - \left[P(b_1+\Delta x) - P(b_2-\Delta x)\right]\right)^j$, where $P(b)$ is the cumulative distribution of $b$;
\item {\it $b_2$ arrives} with probability $p(b_2)db_2$.
\end{enumerate}
To obtain $p_\T{oo}$, we average this over all possible introduction times of $b_1$, sum over the possible values of $j$, and average over all values of $b_1$ and $b_2$ that satisfy conditions in Eq.~(\ref{eq:wrong_order}):
\begin{equation}
\begin{split}
p_\T{oo} =& \frac{1}{N-1}\sum\displaylimits_{t=1}^{N-1}\int_{ } p(b_1)db_1 \!\!\!\! \int\limits_{b_1}^{b_1+\Delta x}\!\!\!\!p(b_2)db_2 \sum\displaylimits_{j=0}^{N-t-1}\left(1 - \left[P(b_1+\Delta x) - P(b_2-\Delta x)\right]\right)^j.
\end{split}
\end{equation}
We evaluate the sums and for simplicity we substitute $b_1\rightarrow b$ and $b_2\rightarrow b+y$, leading to
\begin{equation}\label{eq:poo_exact}
\begin{split}
p_\T{oo} =&\int_{} p(b)db \int\limits_{0}^{\Delta x}\!\!p(b+y)dy\left(\frac{1}{P(b+\Delta x)-P(b+y-\Delta x)} - \right.\\
-&\left.\frac{1-P(b+\Delta x)+P(b+y-\Delta x) - [1-P(b+\Delta x)+P(b+y-\Delta x)]^N}{[N-1][P(b+\Delta x)-P(b+y-\Delta x)]^2}\right).
\end{split}
\end{equation}
Numerically evaluating the above formula and inserting its result into Eq.~(\ref{eq:tauw2}) provides the exact solution of $\tau_\T{tal}(w=2)$, Fig.~\ref{fig:SI-poo}a shows agreement with simulations for various system sizes.

\subsection*{Limit of large $N$ and small $\Delta x$.}
We now calculate $p^{(1)}_\T{oo}$, which is $p_\T{oo}$ in the large population limit such that $\Delta x$ is small.  To proceed, we expand Eq.~(\ref{eq:poo_exact}) with respect to $\Delta x$, assuming $\Delta x\sim N^{-1}$ and ignoring $O(N^{-1})$ and smaller terms, we obtain
\begin{equation}\label{eq:poo_smallDx}
\begin{split}
p^{(1)}_\T{oo}(N,\Delta x) &=\int_{} db \int\limits_{0}^{\Delta x}\!\!dy\left(p(b)\frac{1}{2\Delta x -y} -p^2(b)\frac{1-\exp[-(2\Delta x-y)p(b)N]}{N[(2\Delta x-y)p(b)]^2}\right)=\\
=& \ln 2 - \int_{ }db \!\!\!\!\int\limits_{0}^{N\Delta x p(b)}\!\!\!\!\!\!dY\,p(b)\frac{1-\exp[-(2N\Delta xp(b)-Y)]}{(2N\Delta xp(b)-Y)^2},
\end{split}
\end{equation}
where we relied on the relation $(1-x)^N\approx e^{-xN}$, and in the second integral we made the substitution $y Np(b)\rightarrow Y$. We immediately note the following:
\begin{enumerate}[(i)]
\item The second integral is always positive, therefore $\ln 2$ is the maximum value of $p^{(1)}_\T{oo}$. This maximum is reached in the $N\Delta x\rightarrow\infty$ limit. 
\item $p^{(1)}_\T{oo}(\Delta x, N)$ only depends on  $N\Delta x$, i.e.,  $p^{(1)}_\T{oo}(N,\Delta x)\equiv p^{(1)}_\T{oo}(N\Delta x)$; therefore, in case of small $\Delta x$ , plotting $p_\T{oo}$ as a function of $N\Delta x$ collapses the values of $p_\T{oo}$ for different population sizes $N$ (Fig.~\ref{fig:SI-poo}b). 
\item So far in the calculations, we did not specify $p(b)$; therefore the above properties are true for any continuous unbounded talent distribution.
\end{enumerate}

We now extract the asymptotic behavior of $p_\T{oo}(N\Delta x)$ in the large $N\Delta x$ limit. Preforming the substitution $N\Delta x p(b)\rightarrow P$, we get
\begin{equation}
\begin{split}
p^{(1)}_\T{oo}(N\Delta x) =& \ln 2 - \frac{2}{N\Delta x}\int\displaylimits_0^{\frac{N\Delta x}{\sqrt{2\pi}}} \!\!\!\! dP \int\displaylimits_{0}^{P}\!\!dY\,\frac{P}{\sqrt{2\ln\frac{N\Delta x}{\sqrt{2\pi}}-2\ln P}}\frac{1-\exp[-(2P-Y)]}{(2P-Y)^2}.
\end{split}
\end{equation}
The integrand diverges at both $P=0$ and $P=\frac{N\Delta x}{\sqrt{2\pi}}$; therefore, we separate the first integral in to two sections $[0,\frac{1}{\sqrt{2\pi}}]$ and $[\frac{1}{\sqrt{2\pi}},\frac{N\Delta x}{\sqrt{2\pi}}]$, note that the any $O(1)$ separation point yields the same asymptotic behavior. Investigating the leading order terms in $N\Delta x$, we find that the integral is dominated by the second section. Keeping only the leading order term, we get
\begin{equation}\label{eq:poo_smallDx_asymp}
\begin{split}
p^{(1)}_\T{oo}(N\Delta x) \simeq & \ln 2 - \frac{2}{N\Delta x}\int\displaylimits_{\frac{1}{\sqrt{2\pi}}}^{\frac{N\Delta x}{\sqrt{2\pi}}} \!\!\!\!dP \int\displaylimits_{0}^{P}\!\!dY\,\frac{P}{\sqrt{2\ln\frac{N\Delta x}{\sqrt{2\pi}}-2\ln P}}\frac{1}{(2P-Y)^2}=\\
=& \ln 2 - \sqrt{2}\frac{\sqrt{\ln N\Delta x}}{N\Delta x}.
\end{split}
\end{equation}
Figure~\ref{fig:SI-poo}c compares the exact and asymptotic solution of $p^{(1)}_\T{oo}(N\Delta x)$.

\subsection*{Limit of large $N$ and $\Delta x\sim 1$.}

We turn our attention to calculating $p_\T{oo}^{(2)}$, which is $p_\T{oo}$ in the large population limit such that $\Delta x\sim 1$. Starting from Eq.~(\ref{eq:poo_exact}) and ignoring $O(N^{-1})$ and smaller terms, we immediately obtain
\begin{equation}\label{eq:poo_Dxconst}
\begin{split}
&p^{(2)}_\T{oo}(\Delta x) =\int_{ } p(b)db \int\limits_{0}^{\Delta x}\!\!p(b+y)dy\frac{1}{P(b+\Delta x)-P(b+y-\Delta x)}.
\end{split}
\end{equation}
To extract the asymptotic behavior of $p^{(2)}_\T{oo}(\Delta x)$ in the small $\Delta x$ limit, we expand the cumulative functions in the denominator around $b+y$:
\begin{equation}\label{eq:poo_Dxconst_denom}
\begin{split}
P(b+\Delta x)-&P(b+y-\Delta x)= P(b+y)+p(b+y)(\Delta x -y)+\frac{1}{2}p^\prime(b+y)(\Delta x -y)^2+\ldots \\
 &- P(b+y) + p(b+y)\Delta x -\frac{1}{2}p^\prime(b+y)\Delta x^2 - \ldots \\
 &= p(b+y)(2\Delta x -y)+\frac{1}{2}p^\prime(b+y)\left((\Delta x -y)^2 - \Delta x^2\right)+\ldots.
\end{split}
\end{equation}
First, we focus on the leading order term, substituting back to Eq.~(\ref{eq:poo_Dxconst}):
\begin{equation}\label{eq:poo_Dxconst_limit}
\begin{split}
&p^{(2)}_\T{oo}(N,\Delta x) \xrightarrow[\Delta x\to 0]{}\int_{ } p(b)db \int\limits_{0}^{\Delta x}\!\!dy\frac{1}{2\Delta x -y}=\ln 2.
\end{split}
\end{equation}
Interpreting these results we note the following:
\begin{enumerate}[(i)]
\item Similarly to $p^{(1)}_\T{oo}$, $\ln 2$ is the maximum value of $p^{(2)}_\T{oo}$. This maximum is reached in the $\Delta x\rightarrow 0$ limit.
\item $p^{(2)}_\T{oo}(\Delta x, N)$ does not depend on  $N$, i.e.,  $p^{(2)}_\T{oo}(N,\Delta x)\equiv p^{(2)}_\T{oo}(\Delta x)$; therefore, in case of large $\Delta x$, the values of $p_\T{oo}$ collapse for different system sizes $N$ (Fig.~\ref{fig:SI-poo}d).
\item So far in the calculations, we did not specify $p(b)$; therefore the above properties are true for any continuous unbounded talent distribution.
\end{enumerate}

To extract the small $\Delta x$ behavior of $p^{(2)}_\T{oo}(\Delta x)$, we keep higher order terms of Eq.~(\ref{eq:poo_Dxconst_denom}) and make use of the fact that for standard normal distribution $p^\prime(b) = -bp(b)$, $\avg b=0$, and $\avg{b^2}=1$, thus obtaining
\begin{equation}\label{eq:poo_Dxconst_asymp}
\begin{split}
&p^{(2)}_\T{oo}(\Delta x) = \ln 2 - \left(\ln 2 - \frac{5}{8}\right)\Delta x^2.
\end{split}
\end{equation}
Figure~\ref{fig:SI-poo}e compares the exact and asymptotic solution of $p^{(2)}_\T{oo}(\Delta x)$.

\begin{figure}
	\centering
	\includegraphics[width=1.\textwidth]{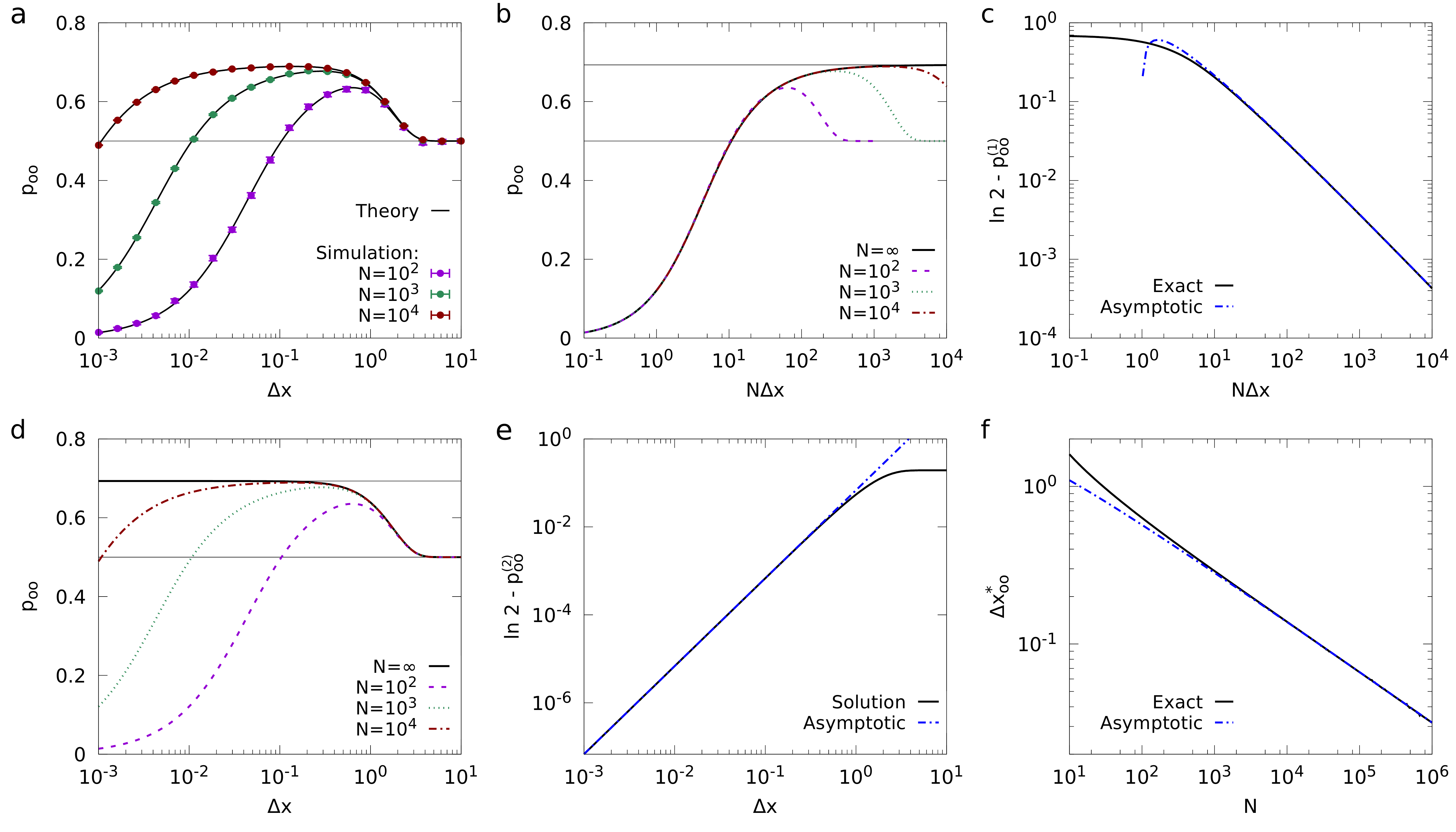}
	\caption{{ \bf Probability of pair in opposite order with respect to talent ($p_\T{\normalfont oo}$).} {\bf(a)}~Comparing the analytical solution of $p_\T{oo}(N,\Delta x)$ (Eq.~(\ref{eq:poo_exact})) to numerical simulations shows excellent agreement for various population sizes. {\bf(b)}~Plotting the exact solution of $p_\T{oo}$ (Eq.~(\ref{eq:poo_exact})) as a function of $N\Delta x$ for various population sizes, the values of $p_\T{oo}$ collapse to $p^{(1)}_\T{oo}(N\Delta x)$ (Eq.~(\ref{eq:poo_smallDx})) for small $N\Delta x$; the range of agreement increases with $N$. {\bf(c)}~Comparing the exact (Eq.~(\ref{eq:poo_smallDx})) and asymptotic (Eq.~(\ref{eq:poo_smallDx_asymp})) solution of $p^{(1)}_\T{oo}(N\Delta x)$. {\bf(d)}~Plotting the exact solution of $p_\T{oo}$ (Eq.~(\ref{eq:poo_exact})) as a function of $\Delta x$ for various system sizes, the values of $p_\T{oo}$ collapse to $p^{(2)}_\T{oo}(\Delta x)$ (Eq.~(\ref{eq:poo_Dxconst})) for large $\Delta x$; the range of agreement increases with $N$. {\bf(e)}~Comparing the exact (Eq.~(\ref{eq:poo_Dxconst})) and asymptotic (Eq.~(\ref{eq:poo_Dxconst_asymp})) solution of $p^{(2)}_\T{oo}(\Delta x)$. {\bf(f)}~Comparing the scaling behavior of $\Delta x^*_\T{oo}$ (Eq.~(\ref{eq:poo_peak})) to results obtained by numerically finding the maximum of the exact $p_\T{oo}(N, \Delta x)$ (Eq.~(\ref{eq:poo_exact})).}
	\label{fig:SI-poo}
\end{figure}

\newpage

\subsection*{Location of maximum of $p_\T{\normalfont oo}(N,\Delta x)$.}

We have calculated $p_\T{oo}(N, \Delta x)$ in two limits: $p^{(1)}_\T{oo}(N \Delta x)$, accurate for small $\Delta x$  (Fig.~\ref{fig:SI-poo}b); and $p^{(2)}_\T{oo}(\Delta x)$, accurate for large $\Delta x$ (Fig.~\ref{fig:SI-poo}d). The function $p^{(1)}_\T{oo}(N \Delta x)$ is monotonic increasing, while function $p^{(2)}_\T{oo}(\Delta x)$ is monotonic decreasing. Therefore the peak of $p_\T{oo}(N, \Delta x)$ is at the crossover between the two limiting cases:
\begin{equation}
\begin{split}
&p^{(1)}_\T{oo}(N \Delta x^*_\T{oo})\sim p^{(2)}_\T{oo}(\Delta x^*_\T{oo}),
\end{split}
\end{equation}
where $\Delta x^*_\T{oo}$ is the location of the maximum of $p_\T{oo}(N, \Delta x)$ for fixed $N$. Using asymptotics from Eqs.~(\ref{eq:poo_smallDx_asymp}) and (\ref{eq:poo_Dxconst_asymp}), we obtain
\begin{equation}\label{eq:poo_peak}
\begin{split}
\frac{\sqrt{\ln N\Delta x^*_\T{oo}}}{N\Delta x^*_\T{oo}}& \sim {\Delta x^*_\T{oo}}^2\quad \rightarrow \quad \Delta x^*_\T{oo}  \sim \frac{(\ln N)^{1/6}}{N^{1/3}}.
\end{split}
\end{equation}
Figure~\ref{fig:SI-poo}f compares the scaling behavior of $\Delta x^*_\T{oo}$ predicted by Eq.~(\ref{eq:poo_peak}) to exact results obtained by numerically finding the maximum of $p_\T{oo}(N, \Delta x)$ provided by Eq.~(\ref{eq:poo_exact}).

\begin{figure}[b]
	\centering
	\includegraphics[width=.4\textwidth]{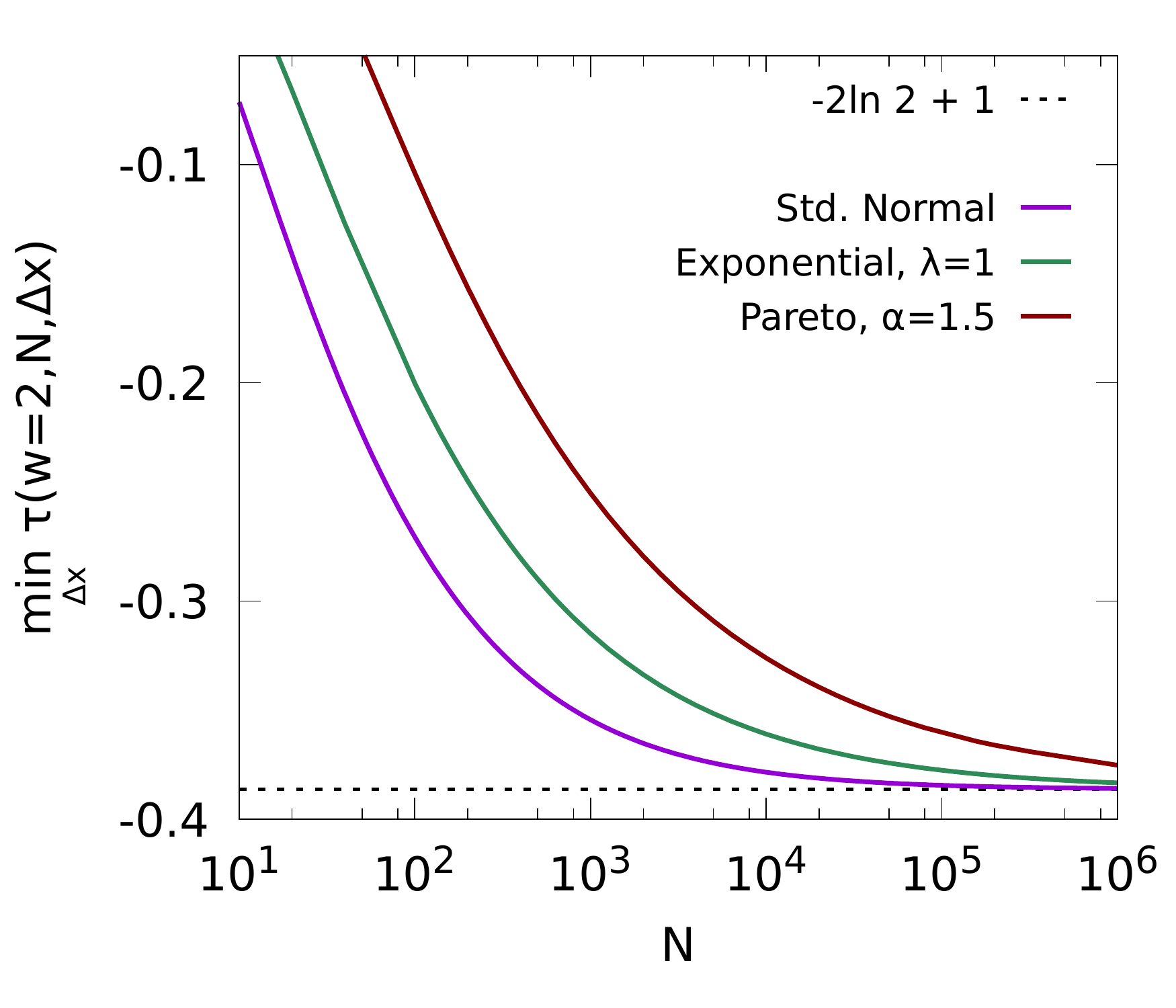}
	\caption{{ \bf Universal minimum of $\tau_\T{\normalfont tal}(w=2)$.} We show $\min\limits_{\Delta x}\tau_\T{tal}(w=2,N,\Delta x)$  numerically calculated using Eq.~(\ref{eq:poo_exact}) for the standard normal distribution, the exponential distribution,  and the Pareto distribution. We find that for fixed $N$, the minimum of $\tau_\T{tal}(w=2)$ depends on the choice of the talent distribution $p(b)$; however, this minimum approaches $-2\ln 2 + 1$ for large populations independent of the details of $p(b)$.}
	\label{fig:SI-taumin}
\end{figure}

\newpage

\subsection*{Universal minimum of $\tau_\T{\normalfont tal}(w=2)$.}
We have shown that $\ln 2$ is the maximum of both $p^{(1)}_\T{oo}(N\Delta x)$ and $p^{(2)}_\T{oo}(\Delta x)$ for any continuous and unbounded $p(b)$.  This means that $\ln 2$ is an upper bound of $p_\T{oo}(N,\Delta x)$, and therefore 
\begin{equation}
\tau_\T{tal}(2)\geq - 2\ln 2 + 1 = -0.386\ldots,
\end{equation}
independent of the choice of $p(b)$ and this minimum is reached in the $N\rightarrow\infty$ limit. On Fig.~\ref{fig:SI-taumin} we show $\min\limits_{\Delta x}\tau_\T{tal}(w=2,N,\Delta x)$  numerically calculated using Eq.~(\ref{eq:poo_exact}) for the standard normal distribution, the exponential distribution $P(b)=\lambda e^{-\lambda b}$ for $\lambda=1$,  and the Pareto distribution $P(b) = 1 - b^{-\alpha}$ ($b>1$) for $\alpha=1.5$. We indeed find that the minimum approaches $-2\ln 2+1$ for all talent distributions.

\section{Probability of removal-induced rank reversals ($p_\T{\normalfont rr}$)}\label{sec:prr}

In this section we calculate $p_\T{rr}$, the probability that the removal of a random individual causes rank reversals. Rank reversal is possible through the following process. Consider three consecutively ranked individuals that have talents $b_1$, $b_2$, and $b_3$, where $b_1$ is ranked highest, and they are in opposite order with respect to talent, i.e., $b_1<b_2<b_3$. If the conditions
\begin{equation}\label{eq:tension_conditions}
\begin{split}
b_2 <& b_1+\Delta x,\\
b_3 <& b_2+\Delta x,\\
b_3 >& b_1+\Delta x\\
\end{split}
\end{equation}
hold, then the ranking is stable. However, if the middle-ranked individual is removed, $b_1$ and $b_3$ switches order. (For brevity, we refer to ``individual with talent $b_i$'' simply as $b_i$.)

To proceed with determining $p_\T{rr}$, we first calculate the probability that $b_1$ is introduced $j+1$ time steps before $b_2$, and $b_3$ is introduced $k+1$ time steps after $b_2$:
\begin{enumerate}
\item {\it $b_1$ arrives} with probability $p(b_1)db_1$;
\item {\it $j$ individuals arrive}, in order for $b_2$ to be consecutive with $b_1$, all $j$ individuals have to either pass $b_1$ or be passed by $b_2$,  the probability of this is $\left(1 - \left[P(b_1+\Delta x)-P(b_2-\Delta x)\right]\right)^j$, where $P(b)$ is the cumulative distribution of $b$;
\item {\it $b_2$ arrives} with probability $p(b_2)db_2$;
\item {\it $k$ individuals arrive}, in order for $b_3$ to be consecutive with $b_2$, all $k$ individuals have to either pass $b_2$ or be passed by $b_3$,  the probability of this is $\left(1 - \left[P(b_2+\Delta x) - P(b_3-\Delta x)\right]\right)^k$;
\item {\it $b_3$ arrives} with probability $p(b_3)db_3$.
\end{enumerate}
To obtain $p_\T{rr}$, we average this over all possible introduction times of $b_2$, sum over the possible values of $k$ and $j$, and average over all values of $b_1$, $b_2$ and $b_3$ that satisfy the conditions in Eq.~(\ref{eq:tension_conditions}):
\begin{equation}\label{eq:prr}
\begin{split}
p_\T{rr}&(N,\Delta x) =\\
 =&\frac{1}{N}\sum\displaylimits_{t=1}^{N}\int_{ } p(b_1)db_1 \int\limits_{b_1}^{b_1+\Delta x}\!\!\!\!p(b_2)db_2 \sum\displaylimits_{k=0}^{N-t-1}\left(1 - \left[P(b_1+\Delta x) - P(b_2-\Delta x)\right]\right)^k \times\\
\times& \int\limits_{b_1+\Delta x}^{b_2+\Delta x}\!\!\!\!p(b_3)db_3\sum\displaylimits_{j=0}^{t-2}\left(1 - \left[P(b_2+\Delta x) - P(b_3-\Delta x)\right]\right)^j.
\end{split}
\end{equation}
We continue following similar steps we used to analyze $p_\T{oo}(N,\Delta x)$ in Sec.~\ref{sec:poo}. We evaluate the sums and for simplicity we substitute $b_1\rightarrow b$, $b_2\rightarrow b+y$, and $b_2\rightarrow b+\Delta x + z$, which leads to
\begin{equation}\label{eq:prr_exact}
\begin{split}
p_\T{rr}&(N,\Delta x) =\frac{1}{N}\int_{ } db \int\limits_{0}^{\Delta x}\!\!dy\int\limits_{0}^{y}\!\!dz\frac{p(b)p(b+y)p(b+\Delta x + z)}{[P_1-P_2][P_3-P_4]} \bigg[N -\frac{1}{P_1-P_2}-\\
-&\frac{1}{P_3-P_4} + \left(\frac{1}{P_1-P_2} +\frac{1-P_3+P_4}{P_3-P_4-P_1+P_2} \right)(1-P_1+P_2)^{N-1} +\\
+& \left(\frac{1}{P_3-P_4} -\frac{1-P_1+P_2}{P_3-P_4-P_1+P_2} \right)(1-P_3+P_4)^{N-1}\bigg],
\end{split}
\end{equation}
where $P_1=P(b+\Delta x)$, $P_2=P(b+y-\Delta x)$, $P_3=P(b+y+\Delta x)$, and $P_4=P(b+z)$. Figure~\ref{fig:SI-prr}a shows the agreement of Eq.~(\ref{eq:prr_exact}) and simulations for various population sizes.

\begin{figure}[t]
	\centering
	\includegraphics[width=1.\textwidth]{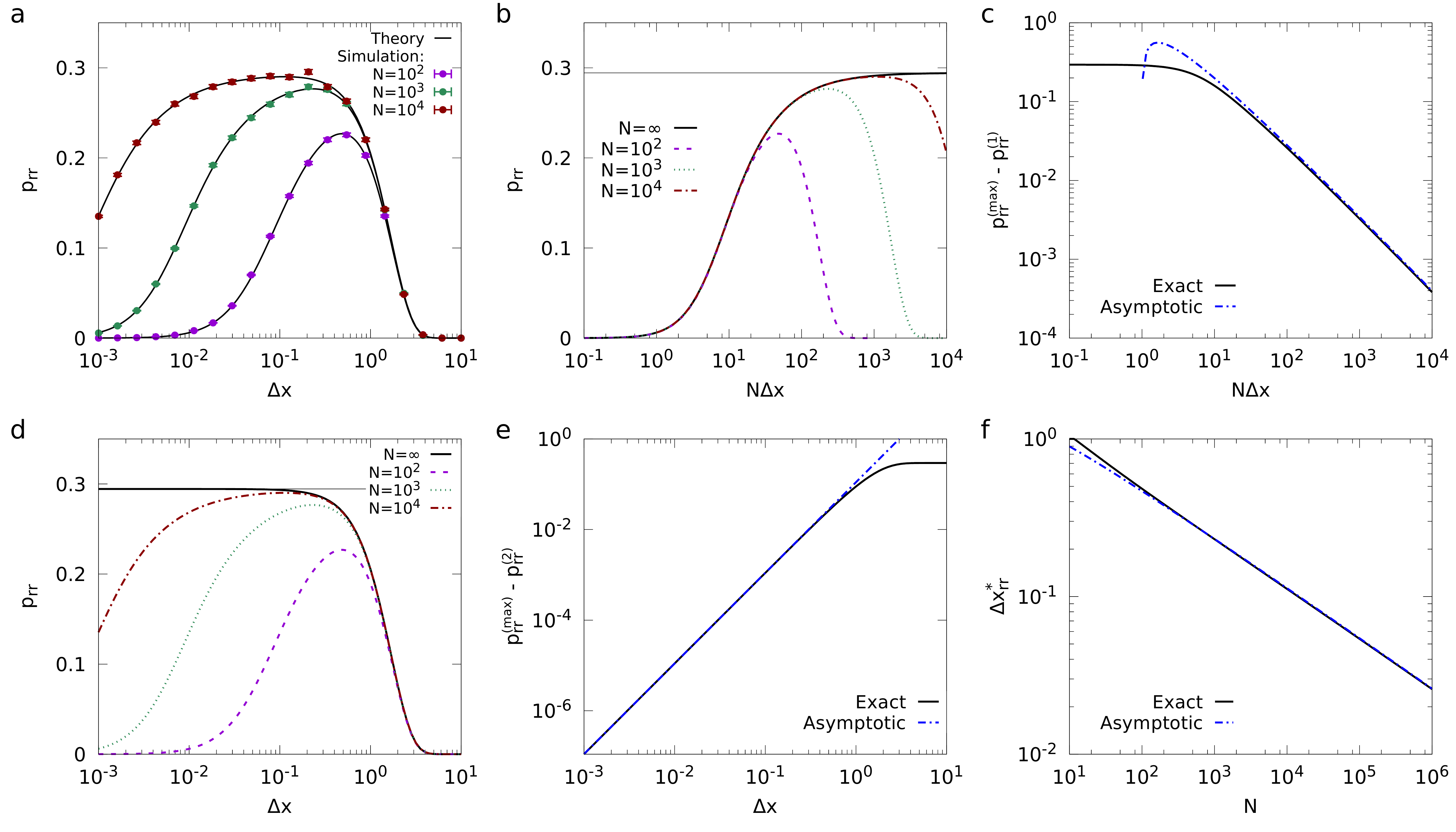}
	\caption{{ \bf Probability that a random removal induces rank reversals ($p_\T{\normalfont rr}$).} {\bf(a)}~Comparing the analytical solution of $p_\T{rr}(N,\Delta x)$ (Eq.~(\ref{eq:prr_exact})) to numerical simulations shows excellent agreement for various population sizes. {\bf(b)}~Plotting the exact solution of $p_\T{rr}$ (Eq.~(\ref{eq:prr_exact})) as a function of $N\Delta x$ for various population sizes, the values of $p_\T{rr}$ collapse to $p^{(1)}_\T{rr}(N\Delta x)$ (Eq.~(\ref{eq:prr_smallDx})) for small $N\Delta x$; the range of agreement increases with $N$. {\bf(c)}~Comparing the exact (Eq.~(\ref{eq:prr_smallDx})) and asymptotic (Eq.~(\ref{eq:prr_smallDx_asymp})) solution of $p^{(1)}_\T{rr}(N\Delta x)$. {\bf(d)}~Plotting the exact solution of $p_\T{rr}$ (Eq.~(\ref{eq:prr_exact})) as a function of $\Delta x$ for various system sizes, the values of $p_\T{rr}$ collapse to $p^{(2)}_\T{rr}(\Delta x)$ (Eq.~(\ref{eq:prr_Dxconst})) for large $\Delta x$; the range of agreement increases with $N$. {\bf(e)}~Comparing the exact (Eq.~(\ref{eq:prr_Dxconst})) and asymptotic (Eq.~(\ref{eq:prr_Dxconst_asymp})) solution of $p^{(2)}_\T{rr}(\Delta x)$. {\bf(f)}~Comparing the scaling behavior of $\Delta x^*_\T{rr}$ (Eq.~(\ref{eq:prr_peak})) to results obtained by numerically finding the maximum of the exact $p_\T{rr}(N, \Delta x)$ (Eq.~(\ref{eq:prr_exact})).}
	\label{fig:SI-prr}
\end{figure}

\newpage

\subsection*{Limit of large $N$ and small $\Delta x$.}
We now calculate $p^{(1)}_\T{rr}$, which is $p_\T{rr}$ in the large population limit such that $\Delta x$ is small.  To proceed, we expand Eq.~(\ref{eq:prr_exact}) with respect to $\Delta x$, assuming $\Delta x\sim N^{-1}$ and ignoring $O(N^{-1})$ and smaller terms, we obtain

\begin{equation}\label{eq:prr_smallDx}
\begin{split}
p^{(1)}_\T{rr}&(N,\Delta x) = \\
\frac{1}{N}&\int_{ } db \int\limits_{0}^{\Delta x}\!\!dy\int\limits_{0}^{y}\!\!dz\frac{1}{(2\Delta x -y)(y+\Delta x -z)} \bigg[ p(b)N-\frac{1}{2\Delta x-y}-\frac{1}{x+\Delta x - y} +\\
+&\left(\frac{1}{2\Delta x -y}+\frac{1}{2y-z-\Delta x}\right)\Exp[-(2\Delta x-y)p(b)N] +\\
+&\left(\frac{1}{y+\Delta x -z}-\frac{1}{2y-z-\Delta x}\right)\Exp[-(y+\Delta x-z)p(b)N]\bigg]=\\
=&p^{(\T{max})}_\T{rr} - \int_{ } db \int\limits_{0}^{\Delta X}\!\!dY\int\limits_{0}^{Y}\!\!dZp(b) \bigg[ \frac{1-\Exp[-(2\Delta X-Y)]}{2\Delta X-Y}+\\
+&\frac{1-\Exp[-(Y+\Delta X-Z)]}{X+\Delta X - Y} +\frac{\Exp[-(2\Delta X-Y)]-\Exp[-(Y+\Delta X-Z)]}{2Y-Z-\Delta X} \bigg],
\end{split}
\end{equation}
where we relied on the relation $(1-x)^N\approx e^{-xN}$; in the second integral we made the substitutions $\Delta xNp(b)\rightarrow \Delta X$, $yNp(b)\rightarrow Y$, and $zNp(b)\rightarrow Z$; and we defined
\begin{equation}\label{eq:prr_UB}
\begin{split}
p^{(\T{max})}_\T{rr} = \ln2\cdot\ln 3 + \T{Li}_2\left(1/3\right)-\T{Li}_2\left(2/3\right)=0.2944\ldots,
\end{split}
\end{equation}
where $\T{Li}_2(x)$ is the second order polylogarithm function. We immediately note the following:
\begin{enumerate}[(i)]
\item The second integral is always positive, therefore $p^{(\T{max})}_\T{rr}$ is the maximum value of $p^{(1)}_\T{rr}$. This maximum is reached in the $N\Delta x\rightarrow\infty$ limit. 
\item $p^{(1)}_\T{rr}(\Delta x, N)$ only depends on  $N\Delta x$, i.e.,  $p^{(1)}_\T{rr}(N,\Delta x)\equiv p^{(1)}_\T{rr}(N\Delta x)$; therefore, in case of small $\Delta x$ , plotting $p_\T{rr}$ as a function of $N\Delta x$ collapses the values of $p_\T{rr}$ for different system sizes $N$ (Fig.~\ref{fig:SI-prr}b). 
\item So far in the calculations, we did not specify $p(b)$; therefore the above properties are true for any continuous unbounded talent distribution.
\end{enumerate}

Following similar steps as in Sec.~\ref{sec:poo}, we obtain the asymptotic behavior for large $N\Delta x$
\begin{equation}\label{eq:prr_smallDx_asymp}
\begin{split}
p^{(1)}_\T{rr}(N\Delta x) \simeq  p^{(\T{max})}_\T{rr} - 2\frac{\sqrt{2}\ln 4}{3}\frac{\sqrt{\ln N\Delta x}}{N\Delta x}.
\end{split}
\end{equation}
Figure~\ref{fig:SI-prr}c compares the exact and asymptotic solution of $p^{(1)}_\T{rr}(N\Delta x)$.

\subsection*{Limit of large $N$ and $\Delta x\sim 1$.}

We turn our attention to calculating $p_\T{rr}^{(2)}$, which is $p_\T{rr}$ in the large population limit such that $\Delta x\sim 1$. Starting from Eq.~(\ref{eq:prr_exact}) and ignoring any $O(N^{-1})$ or smaller term, we immediately obtain
\begin{equation}\label{eq:prr_Dxconst}
\begin{split}
p^{(2)}_\T{rr}(\Delta x) &=\int_{} p(b)db \int\limits_{0}^{\Delta x}\!\!p(b+y)dy\int\limits_{0}^{y}\!\!p(b+ z+\Delta x )dz \,\times \\
\times & \frac{1}{P(b+\Delta x)-P(b+y-\Delta x)}\frac{1}{P(b+y+\Delta x)-P(b+z)}.
\end{split}
\end{equation}
Similarly to Sec.~\ref{sec:poo}, to extract the asymptotic behavior of $p^{(2)}_\T{rr}(\Delta x)$ in the small $\Delta x$ limit, we expand the cumulative functions in the denominator around $b+y$ and $b+z+\Delta x$, respectively. Focusing on the leading order term, we get
\begin{equation}\label{eq:prr_Dxconst_limit}
\begin{split}
&p^{(2)}_\T{rr}(N,\Delta x) \xrightarrow[\Delta x\to 0]{}\int_{} p(b)db \int\limits_{0}^{\Delta x}\!\!dy\int\limits_{0}^{y}\!\!dz\frac{1}{(2\Delta x -y)(y-z+\Delta x)}=p^{(\T{max})}_\T{rr} =0.2944\ldots.
\end{split}
\end{equation}
Interpreting the results we note the following:
\begin{enumerate}[(i)]
\item Similarly to $p^{(1)}_\T{rr}$, $p^{(\T{max})}_\T{rr}$ is the maximum value of $p^{(2)}_\T{rr}$. This maximum is reached in the $\Delta x\rightarrow 0$ limit.
\item $p^{(2)}_\T{rr}(\Delta x, N)$ does not depend on  $N$, i.e.,  $p^{(2)}_\T{rr}(N,\Delta x)\equiv p^{(2)}_\T{rr}(\Delta x)$; therefore, in case of large $\Delta x$, the values of $p_\T{rr}$ collapse for different population sizes $N$ (Fig.~\ref{fig:SI-prr}d).
\item So far in the calculations, we did not specify $p(b)$; therefore the above properties are true for any continuous unbounded talent distribution.
\end{enumerate}

To extract the small $\Delta x$ behavior of $p^{(2)}_\T{rr}(\Delta x)$, we keep higher order terms in the expansion of Eq.~(\ref{eq:prr_Dxconst}) and make use of the fact that for standard normal distribution $p^\prime(b) = -bp(b)$, $\avg b=0$, and $\avg{b^2}=1$:
\begin{equation}\label{eq:prr_Dxconst_asymp}
\begin{split}
p^{(2)}_\T{rr}(\Delta x) = p^{(\T{max})}_\T{rr} - \Big[2+ 3\ln 2 \cdot \ln 3 -4\ln 2 +3 \T{Li}_2(1/3) - 3\T{Li}_2(2/3) \Big]\Delta x^2.
\end{split}
\end{equation}
Figure~\ref{fig:SI-prr}e compares the exact and asymptotic solution of $p^{(2)}_\T{rr}(\Delta x)$.

\subsection*{Location of maximum of $p_\T{\normalfont rr}(N,\Delta x)$.}

We have calculated $p_\T{rr}(N, \Delta x)$ in two limits: $p^{(1)}_\T{rr}(N \Delta x)$, accurate for small $\Delta x$  (Fig.~\ref{fig:SI-prr}b); and $p^{(2)}_\T{rr}(\Delta x)$, accurate for large $\Delta x$ (Fig.~\ref{fig:SI-prr}d). Their scaling properties are identical to $p^{(1)}_\T{oo}(N \Delta x)$ and $p^{(2)}_\T{oo}(\Delta x)$, respectively; therefore, the location of the peak of $p_\T{rr}$ also scales as
\begin{equation}\label{eq:prr_peak}
\begin{split}
\Delta x^*_\T{rr}  \sim \frac{(\ln N)^{1/6}}{N^{1/3}}.
\end{split}
\end{equation}
Figure~\ref{fig:SI-prr}f compares the scaling behavior of $\Delta x^*_\T{rr}$ predicted by Eq.~(\ref{eq:prr_peak}) to exact results obtained by numerically finding the maximum of $p_\T{rr}(N, \Delta x)$ provided by Eq.~(\ref{eq:prr_exact}).

\begin{figure}[b]
	\centering
	\includegraphics[width=.4\textwidth]{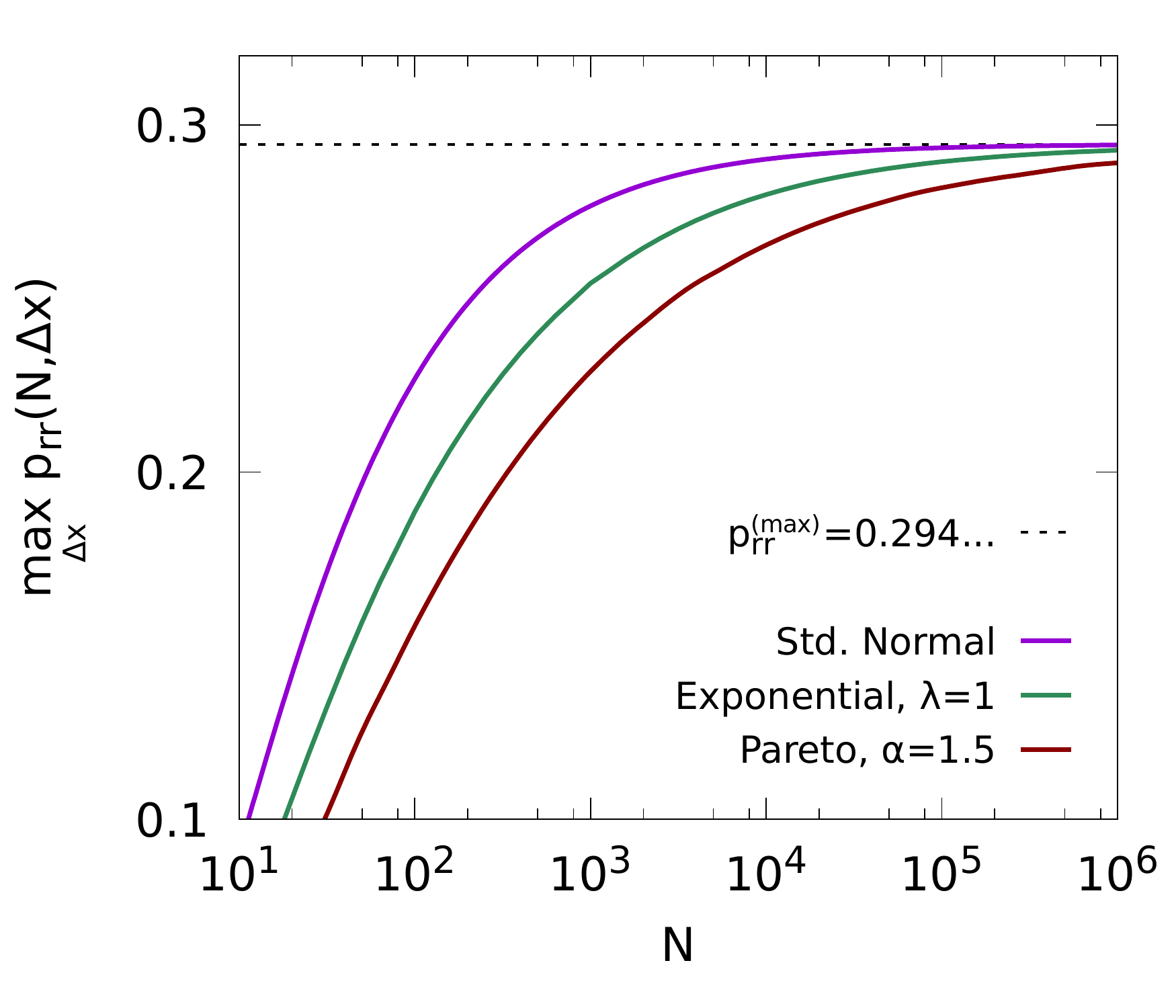}
	\caption{{ \bf Universal maximum of $p_\T{\normalfont rr}$.} We show $\max\limits_{\Delta x}p_\T{rr}(N,\Delta x)$  numerically calculated using Eq.~(\ref{eq:poo_exact}) for the standard normal distribution, the exponential distribution,  and the Pareto distribution. We find that for fixed $N$, the maximum of $p_\T{rr}$ depends on the choice of the talent distribution $p(b)$; however, this maximum approaches $p^{(\T{max})}_\T{rr}=0.294\ldots$ for large populations independent of the details of $p(b)$.}
	\label{fig:SI-prrmax}
\end{figure}

\newpage

\subsection*{Universal maximum of $p_\T{\normalfont oo}(N,\Delta x)$.}
We have shown that $p^{(\T{max})}_\T{rr}$ is the maximum of both $p^{(1)}_\T{rr}(N\Delta x)$ and $p^{(2)}_\T{rr}(\Delta x)$ for any continuous and unbounded $p(b)$.  Therefore $p^{(\T{max})}_\T{rr}$ is an upper bound of $p_\T{rr}(N,\Delta x)$:
\begin{equation}
p_\T{rr}\leq p^{(\T{max})}_\T{rr}=  \ln2\cdot\ln 3 + \T{Li}_2\left(1/3\right)-\T{Li}_2\left(2/3\right)=0.294\ldots,
\end{equation}
independent of the choice of $p(b)$ and this maximum is reached in the $N\rightarrow\infty$ limit. On Fig.~\ref{fig:SI-prrmax} we show $\max\limits_{\Delta x}p_\T{rr}(N,\Delta x)$  numerically calculated using Eq.~(\ref{eq:prr_exact}) for the standard normal distributionthe exponential distribution $P(b)=\lambda e^{-\lambda b}$ for $\lambda=1$,  and the Pareto distribution $P(b) = 1 - b^{-\alpha}$ ($b>1$) for $\alpha=1.5$. We indeed find that the maximum approaches $p^{(\T{max})}_\T{rr}$ for all talent distributions.

\section{Average number of removal-induced rank reversals ($N_\T{\normalfont diff}$)}\label{sec:Ndiff}

In this section, we calculate $N_\T{diff}$, the expected number of pairs that reverse order under the condition that at least one reversal happens. The enumeration of possible events quickly becomes intractable with increasing number of reversals; therefore we turn to an approximate method building on Sec.~\ref{sec:global_corr}. Our approach is based on the observation that the expectation value of the global talent-rank correlation $\tau_\T{tal}$  is stable, and does not fluctuate widely over time. That is the effect of the removal of a random individual and the introduction of a new one at the bottom of the hierarchy on average cancel out each other.

First, we consider the effect of a newly added individual, i.e., an individual added at time step $N$ with talent $b_N$. We calculate the expected number of individuals that are less talented than $b_N$, but are not passed by it, in other words, the number of discordant pairs created by the new individual. The new individual and the one introduced at time step $t$ with talent $b_{t}$ will form  such a discordant pair, if
\begin{equation}
\begin{split}
b_{t}<&b_N,\\
a(N-t,b_{t})>&b_N,
\end{split}
\end{equation}
where $a(N-t,b_{t})$ is the threshold calculated in Eq.~(\ref{eq:a}). Therefore the total number of new discordant pairs is
\begin{equation}
\sum\displaylimits_{t=1}^{N-1} \int_{ } p(b_{t})db_{t}\!\!\!\!\int\limits_{b_{t}}^{a(N-t^\prime,b_{t})}\!\!\!\!\!\!\!\!p(b_N)db_N=N^\T{tal}_-(1),
\end{equation}
where the equality follows from Eq.~(\ref{eq:N-tal}). The function $N^\T{tal}_-(t)$ was introduced in Sec.~\ref{sec:global_corr}, and it  counts the number of individuals that arrive after $t$, have higher talent than the talent of the individual that arrived at $t$, but do not pass it.

Second, we consider the effect of removing a random individual. By removing an individual $i$, discordant pairs are resolved in two ways: (i)~the discordant pairs that $i$ directly participates in are no longer present and (ii)~rank reversals are induced that were previously blocked by $i$. (We are ultimately interested in calculating the number of induced rank reversals.) The total number of discordant pairs is $\sum_{t=1}^N N_-^\T{tal}(t)$, and each pair involves two individuals; therefore the average number of discordant pairs a random individual participates in is
\begin{equation}
\frac{2}{N}\sum_{t=1}^N N_-^\T{tal}(t).
\end{equation}
Since $\tau_\T{tal}$ does not change after removing and adding an individual, the total number of discordant pairs does not change either, mathematically this means
\begin{equation}\label{eq:Ndiff}
N^\T{tal}_-(1) - \frac{2}{N}\sum_{t=1}^N N_-^\T{tal}(t) - N_\T{diff}^{(0)}=0, 
\end{equation}
where the first term is the effect of adding a new individual (the number of newly created discordant pairs), the second and third terms are the number of discordant pairs resolved. From this we calculate $N_\T{diff}^{(0)}$, and Fig.~\ref{fig:SI-Ndiff}a compares simulations to the analytical predictions, finding excellent agreement for various population sizes.

Finally, $N_\T{diff}^{(0)}$ counts the events, including when no rank reversals are induced. In the main text Fig.~3, we show the expected number of pairs that reverse order under the condition that at least one reversal happens, this can be calculated as
\begin{equation}
N_\T{diff}=\frac{N_\T{diff}^{(0)}}{p_\T{rr}}=\frac{N^\T{tal}_-(1) - \frac{2}{N}\sum_{t=1}^N N_-^\T{tal}(t)}{p_\T{rr}},
\end{equation}
where $p_\T{rr}$ is the probability that at least one rank reversal happens, and it is provided by Eq.~(\ref{eq:prr}). Note that $N_\T{diff}^{(0)}$ is related to global correlations, while the probability that a rank reversal happens $p_\T{rr}$ is a local property dependent on the relation of only three consecutive individuals.

\subsection*{Continuous time approximation.}

To determine the scaling behavior of $N^{(0)}_\T{diff}(\Delta x, N)$, we make use of the continuous time approximation introduced in Sec.~\ref{sec:global_corr}. Following Eq.~(\ref{eq:Ntal_ct}), we can write
\begin{equation}\label{eq:Ndiff_components_ct}
\begin{split}
N_-^\T{tal}(1) &= N\int_{0}^{1}\!\!ds \!\!\int_{ } db p(b)P(a(s\cdot\Delta x^2N,b))-\frac{N}{2},\\
\frac{2}{N}\sum\limits_{t=1}^N N_-^\T{tal}(t) &= 2N\int_{0}^{1}\!\!ds \!\!\int_{ } db p(b) (1-s) P(a(s\cdot\Delta x^2N,b))-\frac{N}{2}.
\end{split}
\end{equation}
Therefore, using Eq.~(\ref{eq:Ndiff}), the average number of rank reversals in the continuous time approximation is
\begin{equation}\label{eq:Ndiff_ct}
\begin{split}
N^{(0)}_\T{diff}(\Delta x, N) &= N\int_{0}^{1}\!\!ds \!\!\int_{ } db p(b) (2s-1) P(a(s\cdot\Delta x^2N,b)),
\end{split}
\end{equation}
this result is accurate for large populations. We can now see that
\begin{enumerate}[(i)]
\item The average number of rank reversals scale as $N^{(0)}_\T{diff}(\Delta x, N)\equiv Nf(\Delta x^2 N)$ for large $N$, where $f(x)$ is the sclaing function defined by Eq.~(\ref{eq:Ndiff_ct}); therefore plotting $N^{(0)}_\T{diff}/N$ as a function of $\Delta x \sqrt{N}$ collapses the values of $N^{(0)}_\T{diff}$ for large populations (Fig.~\ref{fig:SI-Ndiff}b).
\item Therefore the point $\Delta x^*_\T{diff}$ where $N^{(0)}_\T{diff}$ reaches its maximum, scales as $\Delta x^*_\T{diff}\sim 1/\sqrt{N}$, the same as the crossover point $\Delta x_\T c$ (Fig.~\ref{fig:SI-Ndiff}c).
\item If $\Delta x^2 N\rightarrow 0$ then $P(a(s\cdot\Delta x^2N,b))\rightarrow 0$ and if $\Delta x^2 N\rightarrow \infty$ then $P(a(s\cdot\Delta x^2N,b))\rightarrow 1$ (for $s>0$); therefore in both cases the continuous time approximation predicts $N^{(0)}_\T{diff}\rightarrow 0$.
\item So far in the calculations, we did not specify $p(b)$; therefore the above properties are true for any continuous unbounded talent distribution.
\end{enumerate}

\begin{figure}
	\centering
	\includegraphics[width=1.\textwidth]{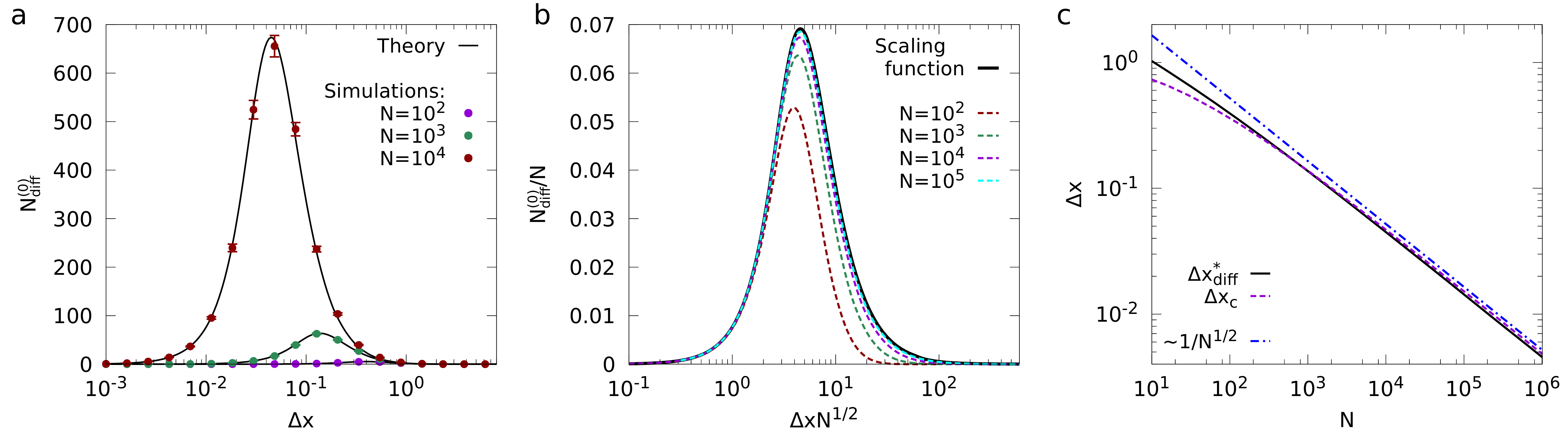}
	\caption{{ \bf Average number of rank reversals $N_\T{\normalfont diff}^{(0)}$.} {\bf(a)}~Comparing the discrete time analytical solution of $N_\T{diff}^{(0)}$ (Eqs.~(\ref{eq:Ndiff}) to numerical simulations shows excellent agreement for various system sizes. Error bars represent the standard error of the mean.  {\bf(b)}~The continuous time approximation predicts that plotting $N_\T{diff}^{(0)}/N$ as a function of $\Delta x \sqrt{N}$ collapses the values of $N_\T{diff}^{(0)}$. We compare the rescaled discrete time solution of $N_\T{diff}^{(0)}$ to the scaling function, provided by Eq.~(\ref{eq:Ndiff_ct}). We find that the scaling is accurate for large populations. {\bf(c)}~We obtain $\Delta x^*_\T{diff}$, the location of the peak of $N_\T{diff}^{(0)}$, by numerically finding the maximum of Eq.~(\ref{eq:Ndiff}). The location of the peak scales as $\Delta x^*_\T{diff}\sim 1/\sqrt{N}$ for large populations in accordance to the continuous time approximation. We find that $\Delta x^*_\T{diff}$ is closely related to the crossover point $\Delta x_\T{c}$.}
	\label{fig:SI-Ndiff}
\end{figure}

\section{Properties of the hierarchy in the $N\rightarrow\infty$ limit}
\label{sec:Ninfty}

In the previous sections, we have showed that local properties (local correlation $\tau_\T{tal}(2)$ and probability of rank reversals $p_\T{rr}$) and global properties (global correlations $\tau_\T{tal}$, $\tau_\T{exp}$ and the average number of rank reversals $N^{(0)}_\T{diff}$) have different scaling properties. Therefore the emergent hierarchy may posses different properties in the large population limit $N\rightarrow \infty$ depending on the relationship between $N$ and $\Delta x$. Here we explore possible outcomes assuming
\begin{equation}\label{eq:alpha_def}
N^\alpha \Delta x = C,
\end{equation}
where $C$ is constant.

Note that in the parametrization of the model, provided in Eq.~(\ref{eq:cond_to_pass_A}), $\Delta x=\delta/[\mu(N-1)]$, i.e., $\alpha=1$. Other values of $\alpha$ are also possible through adjustment of $\delta$ or $\mu$, or if individuals do not randomly select opponents, but selectively compete with similarly ranked ones.

\renewcommand{\arraystretch}{1.2}
\begin{table}[b]
\centering
\begin{tabular}{c|x{2.5cm}x{2.5cm}x{2.5cm}x{2.5cm}x{2.5cm}}
& \multicolumn{5}{ c }{In the limit of $N\rightarrow\infty$} \tabularnewline[-2ex]
& $\tau_\T{tal}$ & $\tau_\T{exp}$ &  $\tau(2)$ & $p_\T{rr}$ & $N^{(0)}_\T{diff}/N$ \tabularnewline
\hline\hline
$1<\alpha$ & 1 & 0 & 1 & 0 & 0 \tabularnewline
\hline
$\alpha=1$ & 1 & 0 & $1-2p^{(1)}_\T{oo}(C)$ & $p^{(1)}_\T{rr}(C)$ & 0 \tabularnewline
\hline
$1/2<\alpha<1$ & 1 &  0 & $1-2\ln 2$ & $0.2944\ldots$ & 0 \tabularnewline
\hline
$\alpha=1/2$ & $\tau_\T{tal}(C)$ & $1-\tau_\T{tal}(C)$ & $1-2\ln 2$ & $0.2944\ldots$ & $f^
{(0)}(C)$ \tabularnewline
\hline
$0<\alpha<1/2$ & 0 & 1 & $1-2\ln 2$ & $0.2944\ldots$ & 0 \tabularnewline
\hline
$0=\alpha$ & 0 & 1 & $1-2p^{(2)}_\T{oo}(C)$ & $p^{(2)}_\T{rr}(C)$ & 0 \tabularnewline
\hline
$\alpha<0$  & 0 & 1 & 0 & 0 & 0
\end{tabular}
\caption{\label{tab:limits} {\bf Structure of hierarchy in the large population limit.} The numerical values are valid for any continuous unbounded talent distribution, while the scaling functions depend on the particular choice. The scaling functions are defined by the corresponding equations: $\tau_\T{tal}(C)$ in Eq.~(\ref{eq:tau_b_ct}), $p^{(1)}_\T{oo}(C)$ in Eq.~(\ref{eq:poo_smallDx}),  $p^{(2)}_\T{oo}(C)$ in Eq.~(\ref{eq:poo_Dxconst}), $p^{(1)}_\T{rr}(C)$ in Eq.~(\ref{eq:prr_smallDx}),  $p^{(2)}_\T{rr}(C)$ in Eq.~(\ref{eq:prr_Dxconst}), and $f^{(0)}(C)$ in Eq.~(\ref{eq:Ndiff_ct}).}
\end{table}

Table~\ref{tab:limits} enumerates the possible emergent hierarchies in the large population limit as a function of $\alpha$. The most rich behavior is exhibited by the $\alpha = 1/2$ case, characterized by both nonzero  global correlation with  both talent and experience, maximal local anti-correlation and rank reversal probability, and large-scale rank rearrangements. Other counterintuitive scenarios are also possible. For example, for $0<\alpha<1/2$, despite dynamics that aim to order the hierarchy according to talent, global talent correlation is zero and local correlation is negative. We anticipate that talent correlation is positive on an intermediate scale. Finally, note that negative local correlation and non-zero rank reversal probability are robust features of the hierarchies, $\tau_\T{tal}(2)$ and $p_\T{rr}$ become trivial only for the unrealistic cases of $1<\alpha$ and $\alpha<0$.

